\documentclass[12pt]{article}
\usepackage[letterpaper,margin=1in]{geometry}

\usepackage{pdfpages} 
\usepackage{standalone} 

\usepackage{booktabs}
\usepackage{xcolor}
\usepackage{xr}
\usepackage{multirow}%
\usepackage{amsmath,amssymb,amsfonts}%
\usepackage{amsthm}%
\usepackage{mathrsfs}%
\usepackage[title]{appendix}%
\usepackage{xcolor}%
\usepackage{textcomp}%
\usepackage{manyfoot}%
\usepackage{booktabs}%
\usepackage{algorithm}%
\usepackage{algorithmicx}%
\usepackage{algpseudocode}%
\usepackage{listings}%
\theoremstyle{thmstyleone}%
\theoremstyle{thmstyletwo}%
\theoremstyle{thmstylethree}%
\usepackage{graphicx}
\usepackage{epstopdf} 

\renewenvironment{abstract}
	{\quotation}
	{\endquotation}

\date{}

\usepackage{scicite}

\title{Reconfigurable Physical Unclonable Function based on SOT-MRAM Chips}

\author{
	Min~Wang$^{1, \dagger}$,
    Chuanpeng~Jiang$^{1, \dagger}$,
    Zhaohao~Wang$^{1, 2, \ast}$,
 	Zhengyi~Hou$^{1}$,\\
    Zhongkui~Zhang$^{1}$,
    Yuanfu~Zhao$^{3, \ast}$,
    Hongxi~Liu$^{4, \ast}$,
	Weisheng~Zhao$^{1, \ast}$\and
	\small$^{1}$~School of Integrated Circuit Science and Engineering, \\
    \small Beihang University, Beijing \& 100191, China.\and
	\small$^{2}$~National Key Laboratory of Spintronics, Hangzhou International Innovation Institute, \\
    \small Beihang University, Hangzhou \& 311115, China.\and
    \small$^{3}$~Beijing Microelectronics Technology Institute, Beijing \& 100076, China.\and
    \small$^{4}$~ Truth Memory Corporation, Beijing \& 100088, China.\and
    \small$^\dagger$These authors contributed equally to this work.\and	
	\small$^\ast$Corresponding authors. Email: zhaohao.wang@buaa.edu.cn, zhaoyf@vip.163.com, \and 
    \small hongxi\_liu@tmc-bj.cn, weisheng.zhao@buaa.edu.cn
}

\begin{document}

\maketitle

\begin{abstract} \bfseries \boldmath
Hardware-based security primitives have become critical to enhancing information security in the Internet of Things (IoT) era. Physical unclonable functions (PUFs) utilize the inherent variations in the manufacturing process to generate cryptographic keys unique to a device. Reconfigurable PUFs (rPUFs) can update cryptographic keys for enhanced security in dynamic operational scenarios involving huge amounts of data, which makes them suitable for implementation in CMOS-integrated spin-orbit torque magnetic random access memory (SOT-MRAM) chips. However, a key challenge is achieving real-time reconfiguration independent of the environmental conditions, particularly the operating temperature. We propose a dual-pulse reconfiguration strategy for rPUFs in CMOS-integrated SOT-MRAM chips that effectively widens the operating window and achieves resilience across a wide range of operating temperatures without the need for dynamic feedback that overly complicates circuit design. The proposed strategy lays a solid foundation for the next generation of hardware-based security primitives to protect IoT architectures.
\end{abstract}

\section{Introduction}

Information security requires multilayered encryption spanning both software and hardware. The expansion of the Internet of Things (IoT) has made millions of edge devices and systems vulnerable to cyberthreats, which has made information security at the hardware level increasingly critical \cite{japa2021hardware, ghosh2016spintronics,hu2020overview}. Key hardware-based security primitives include the Trusted Platform Module, Hardware Security Module, and Physical Unclonable Function (PUF). In particular, PUFs are a lightweight and cost-effective solution that generate cryptographic keys specific to a chip based on inherent variations in the manufacturing process \cite{herder2014physical, al2023physical, wali2023hardware, chang2017retrospective}. These cryptographic keys are in the form of challenge–response pairs (CRPs), which serve as a digital fingerprint and guarantee the uniqueness and non-clonability of the chip.

PUF designs based on complementary metal–oxide–semiconductor (CMOS) technology feature circuit compatibility and employ inherent variations in the manufacturing process as a static entropy source  \cite{satpathy2019all,taneja2021memory,song2021environmental, park2022ber, he2021AnAutomatic,cortez2015intelligent}. To satisfy the requirements of high-frequency interaction scenarios, stronger PUF designs are being investigated that expand the CRP space exponentially. However, once the cryptographic key is compromised, fixed CRPs are at risk of being permanently invalidated \cite{kraleva2023cryptanalysis}. The above limitations have given rise to the reconfigurable PUF (rPUF), which is an emerging security primitive that utilizes dynamic entropy sources to update the CRP space. This not only expands the CRP space but also enhances security and flexibility against machine learning attacks and untrusted manufacturing risk \cite{john2021halide, gao2022unified, Zhang2022Reconfigurable}. Nonetheless, CMOS-based PUFs hold weak cycle-to-cycle (C2C) variability and cannot support higher-order reconfiguration counts.

Advances in non-volatile memory (NVM) have introduced novel entropy sources that allow rPUFs to be implemented with enhanced security metrics \cite{ding2021unified, sahay2017recent,mahmoodi2019experimental, chen2016review, lim2023dual, park2025physical} in various memory technologies, which include the ferroelectric field-effect transistor (FeFET) \cite{shao2023novel, li2025demonstration, guo2021exploiting, shao2024imce}, phase-change random access memory (PC-RAM) \cite{Le2014Exploiting}, resistance random access memory (RRAM)  \cite{Pang2019Areconfigurable, gao2022unified, Zhao2020AFULLY,cao202467f}, spin-transfer torque magnetic random access memory (STT-MRAM) \cite{wang2017hardware,chen2017hardware,du2023intrinsic}, and spin-orbit torque magnetic random access memory (SOT-MRAM) \cite{Zhang2022Reconfigurable,cao2021reconfigurable, Lee2022Spintronics,zhao2023purely}. For NVM-based PUFs, the balance between endurance and reliability is an important consideration. SOT-MRAM is considered an ideal replacement for static random access memory because of its high reliability, high speed, and low power consumption\cite{dieny2020opportunities, guo2021spintronics}. SOT-MRAM stores binary data in a magnetic tunnel junction (MTJ), which provides inherent dynamic entropy for excellent C2C variability and a physical basis for key regeneration \cite{yin2022scalable,carboni2019stochastic} as well as unlimited endurance for high-order reconfiguration counts \cite{xu2023full}. Thus, SOT-MRAM is an ideal carrier for multifunctional PUFs, which are considered a prime candidate for the next generation of hardware-based security primitives.

Previous studies on rPUFs for SOT-MRAM chips have generally adopted C2C variability with stochastic characteristics based on perpendicular and in-plane anisotropic MTJs, such as variations induced by a switching current ($I_c$) \cite{Zhang2022Reconfigurable}, domain wall nonlinear dynamics \cite{Zhang2020spin}, chiral domain wall motion \cite{cao2021reconfigurable}, and random switching enhanced by self-write-back (SWB) \cite{Koh2023Improved}. Unfortunately, these designs remain at the device or small-scale array level, and practical applications will inevitably require integrated control circuits such as CMOS modules. There is still a lack of analysis on the chip-level reconfiguration and external environment robustness of CMOS-integrated SOT-MRAM chips, for which temperature variations can cause shifts in both the transistor and MTJ parameters. 

Two main metrics are affected by the temperature: the read reliability and real-time reconfiguration. Various studies on PUFs have focused on enhancing the read reliability over a wide temperature range, such as by SWB \cite{song2021environmental, Koh2023Improved}, temporal majority voting (TMV) \cite{gao2022concealable,park2022ber,zhao20201A1036}, unsteady bit masking or discard \cite{park2022ber,he2021AnAutomatic,zhao20201A1036,gao2022concealable}, and feedback compensation \cite{zhao20201A1036,taneja2018fully,brussenskiy2015robust,cortez2015intelligent}. Meanwhile, fewer studies have considered ensuring that real-time reconfiguration can be performed over a wide temperature range. Real-time reconfiguration requires a stable operating window, and a major challenge is the temperature-dependent drift of the critical current for the MTJ. Moreover, conventional feedback compensation techniques are not applicable to CMOS-integrated SOT-MRAM chips, so new schemes for rPUFs need to be designed. However, the physical mechanism by which the temperature affects CMOS-integrated SOT-MRAM chips is still unclear, and the real-time reconfiguration against external temperature fluctuations faces two major bottlenecks: (i) a narrow operating window that requires highly precise circuit design; (ii) the shift in dynamic probability with the operating temperature that increases the complexity of the compensation circuit.

In this work, we propose and experimentally demonstrate a temperature-resilient rPUF design in CMOS-integrated SOT-MRAM chips. The novel dual-pulse strategy increases the operating window for real-time reconfiguration and compensates for shifts in the operating temperature. The proposed rPUF design is a low-cost and streamlined solution for enhanced robustness against temperature with high potential for lightweight hardware-based security primitives.


\section{Results}

\begin{figure}
 \includegraphics[width=\linewidth]{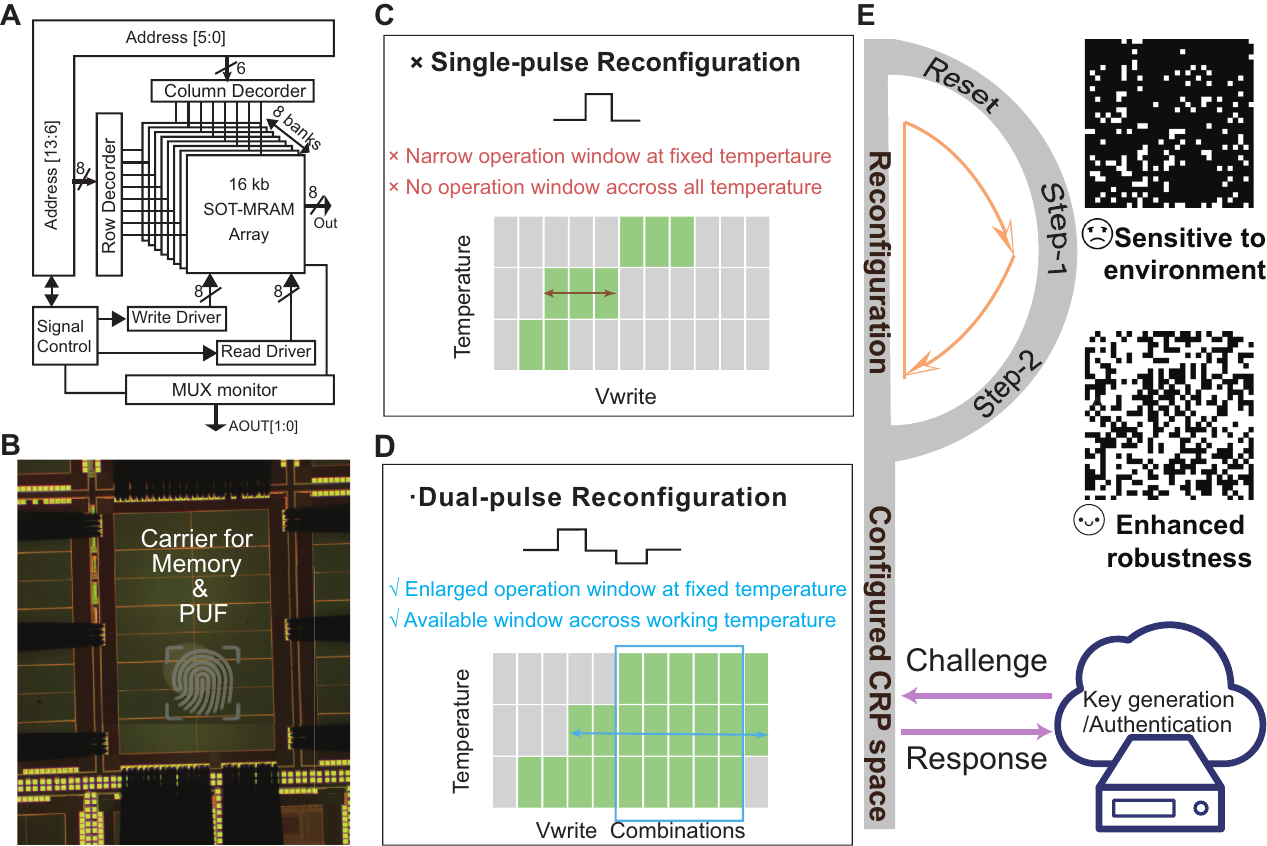}
 \caption{\textbf{Overview of the proposed rPUF.} (\textbf{A}) Architecture of a 128 kb SOT-MRAM chip comprising eight banks. (\textbf{B}) Optical microscopy image of the SOT-MRAM chip during the chip-probing test. (\textbf{C}) Single-pulse reconfiguration strategy. (\textbf{D}) Proposed dual-pulse reconfiguration strategy. (\textbf{E}) Workflow of the proposed rPUF and its authentication protocol. The inserted binary bitmaps indicate the generated PUF responses after the first and second set pulses.}
 \label{fig1overall}
\end{figure}

Figure \ref{fig1overall}A illustrates the architecture of a 128 kb SOT-MRAM chip that was fabricated by using 180 nm CMOS technology on an 8-inch platform \cite{jiang2024demonstration} and integrates eight banks of 16 kb SOT arrays, a read circuit, a write driver, decoders, and buffers. Notably, the multiplexer module is designed to allow access to bit cells in analog mode. In other words, bit cells can flexibly operate with either digital or analog signals depending on whether the multiplexer module is disabled or enabled. One bit of data is stored and read out from a pair of MTJs; in memory mode or after SWB operation, the data are stored in complementary pairs. This pair design contributes to differential read circuits and high read reliability, while also preventing side-channel attacks because of the weak fluctuations in the current when 0 or 1 is read. This type of MTJ has in-plane magnetic anisotropy, which has been demonstrated to be a reliable carrier because of its all-electric write operation and high back-end-of-line compatibility. 

The SOT-MRAM chip can serve as a shared carrier for both memory and PUF modes. The functionality of the chip was verified by writing a shmoo plot (Fig. \ref{figSI:Shmoo}), and we verified memory mode in previous studies \cite{jiang2024demonstration,xiong2024first}. Regarding PUF mode, the SOT-MRAM chip features both static entropy and dynamic entropy that are induced by the fabrication process and thermal noise, respectively. Some of the metrics used to evaluate memory mode can also be implemented in PUF mode. For example, retention of over 10 years indicates long-term reliability in PUF mode. The endurance was found to be more than $1\times 10^{14}$ \cite{xu2023full}, which ensures high-order reconfiguration cycles.

In this study, experiments were conducted by using the chip-probing test, as shown in Fig. \ref{fig1overall}B. The write pulse width was set to 20 ns by default. The analog voltage generated by the chip-probing test machine was applied through the multiplexer module as the write voltage ($V_{write}$). Figure \ref{fig1overall}C shows the conventional one-pulse reconfiguration strategy for PUF reconfiguration, which comprises reset and one-step set pulses. The green regions correspond to the available write voltage for PUF reconfiguration according to temperature. Because of the influence of the thermal disturbance field, the write success rate (WSR) is a sigmoid-like and monotonically increasing function of the write voltage, which inherently limits the width of the operating window at a fixed temperature. Because the WSR depends on the temperature, the narrow operating window results in no crossover between different temperatures. No universal operating window exists across the temperature range of --40°C to 125°C. 

Because the SOT-MRAM chip may undergo reconfiguration at any moment, it is not possible to wait for the temperature to return to room temperature. Therefore, the one-pulse reconfiguration strategy is difficult to integrate with conventional temperature feedback techniques. Figure \ref{fig1overall}D shows our proposed dual-pulse reconfiguration strategy for widening the operating window and enhancing robustness against temperature. This strategy employs two pulses of opposite polarities and decreasing amplitudes, which modifies the monotonically increasing probability curve of the conventional one-pulse reconfiguration strategy to widen the operating window both at a fixed temperature and across the entire temperature range of --40°C to 125°C.

Figure \ref{fig1overall}E illustrates the workflow of the rPUF in an SOT-MRAM chip. First, the chip is initialized to 1 values (i.e., FF polarized writing). Next, the first set pulse of 00 polarity is applied to form a map of stochastic states, which may be affected by the operating temperature and deviate from the ideal distribution. Finally, a second set pulse of FF polarity is applied to correct the data distribution. The above process is denoted as FF–00–FF polarized writing, while the reverse is denoted as 00–FF–00 polarized writing. The related authentication protocol is as follows. When the response needs to be refreshed, the reconfiguration starts, and the response stored by the SOT-MRAM chip is written randomly via the dual-pulse reconfiguration strategy. After reconfiguration, the SOT-MRAM chip stores stable responses and interacts with external devices for authentication in the form of a CRP. Chips from the same manufacturer or after a new reconfiguration can vary greatly in CRP space, which greatly reduces the risk of being copied or predicted and ensures system security. To distinguish between single-bit and chip-level statistics, we employed three complementary metrics: the Hamming weight (HW) to quantify the distribution of 1 values at the chip level, the WSR compared to the initialization state, and the switching probability $P_{sw}$ to characterize single-bit probabilistic behavior. In the single-pulse reconfiguration strategy, the HW directly corresponds to the WSR during 00–FF processes. For FF–00 processes, HW = 1-WSR. The bit error rate (BER) was used as a metric to evaluate the read reliability.

\subsection{Experimental Validation at Room Temperature}

\begin{figure}[!th]
 \includegraphics[width=\linewidth]{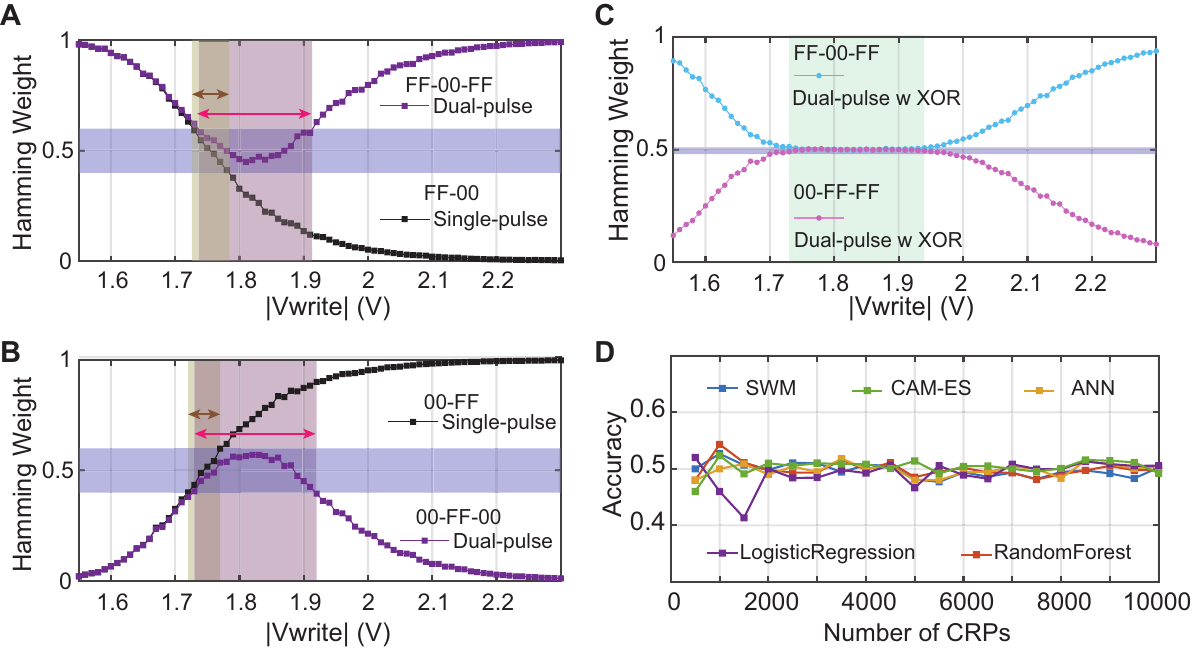}
 \caption{ \textbf{PUF performance with the dual-pulse reconfiguration strategy.} Hamming weight (HW) as a function of the write voltage with different reconfiguration strategies: (\textbf{A}) FF–00 (single-pulse) and FF–00–FF (dual-pulse) polarized writing and (\textbf{B}) 00–FF (single-pulse) and 00–FF–00 (dual-pulse) polarized writing. The target operating window is indicated in blue. $\beta$ = 0.15 V was used. (\textbf{C}) Enhancement of the HW with XOR postprocessing. (\textbf{D}) Robustness against machine learning attacks. The prediction accuracy of five machine learning models was $\sim$50\%.}
 \label{fig:unif}
\end{figure}

We experimentally validated the proposed rPUF by first evaluating the operating window at room temperature. Figure \ref{fig:unif}A shows the measured HW according to the reconfiguration strategy with FF–00 (single-pulse) and FF–00–FF(dual-pulse) polarized writing. With the single-pulse reconfiguration strategy, the HW increased monotonically with the write voltage, similar to a sigmoid-like function. With the dual-pulse reconfiguration strategy, the HW approximated a parabolic function because of the correction by the second set pulse. The first and second set pulses held different polarities, and the relationship between the amplitudes of the first set (V1) and second set (V2) can be defined as
\begin{equation}
 |V2| = |V1| - \beta
 \label{eq:V1V2}
\end{equation}
where $\beta$ is a correction coefficient. As shown in Fig. \ref{fig:unif}B, the same behavior was observed for 00–FF and 00–FF–00 polarized writing. The preliminary target operating window (blue color) limited the HW to 0.4–0.6, which is a prerequisite for XOR postprocessing (see Section \ref{secXOR}). The XOR postprocessing obfuscates PUF data by addressing spatial correlations from process deviations and enhancing key metrics. Figure \ref{fig:unif}C shows the ideal HW and large operating window obtained with XOR postprocessing. Thus, the proposed dual-pulse reconfiguration strategy facilitated a wider operating window while the generated PUF responses retained excellent unpredictability. To evaluate the ability of PUF against machine learning attacks, five classical algorithms were modeled and executed (details in Sec. \ref{SecMLattacks})\cite{wang2020lattice, gu2020modeling, cao2018compact,ebrahimabadi2021puf,chen2017hardware,ashtari2019new, li2023reconfigurable}. The raw prediction accuracy was 55.29\% without postprocessing (Fig. \ref{figSI:ML}), which confirmed the entropy from C2C and device-to-device (D2D) variability. As shown in Fig. \ref{fig:unif}D, a nearly ideal random guessing level (50.03\%) was achieved after XOR postprocessing, which confirmed the robustness of the rPUF against machine learning attacks. 

In terms of uniformity, the dual-pulse reconfiguration strategy combined with XOR postprocessing led to a near-ideal Gauss distribution of rPUF responses with 0.5001 $\mu$ and 0.042 $\sigma$, where the fitted distribution was tightly clustered around $\mu$ (Fig. \ref{figSI:Uniformity_002FF200} and \ref{figSI:Uniformity_FF2002FF}). The randomness and unpredictability of the rPUF responses were further demonstrated by the results of the Automatic Correction Function test (Fig. \ref{figSI:ACF}) and NIST SP800-22 test (Table \ref{Table:NIST}). The enhanced reliability of our design was also explored in terms of the TMV and SWB operations. TMV is an error correction technique that performs multiple read operations in a selected address and uses most of the readout data as the final data. TMV is suitable for energy-saving scenarios because the reading process consumes less energy than the writing process. SWB rewrites the memory cell deterministically according to the readout data, which mitigates unstable bits and optimizes reliability metrics such as the intra-Hamming distance (intra-HD) and BER. The inter-/intra-HD ratio was approximately infinite at ~1500 (Fig. \ref{figSI:Intra_Inter}A, \ref{figSI:Intra_Inter}B), and the average BER was $3.29\times10^{-5}$. In addition, the inter-die HD ($\mu$=0.5009, $\sigma$=0.0437) demonstrated the excellent uniqueness of the proposed rPUF (Fig. \ref{figSI:Intra_Inter}B).

\subsection{Demonstration of Reconfigurability \label{Sec:Reconfiguration}}

\begin{figure}[!h]
\centering
\includegraphics[width=\linewidth]{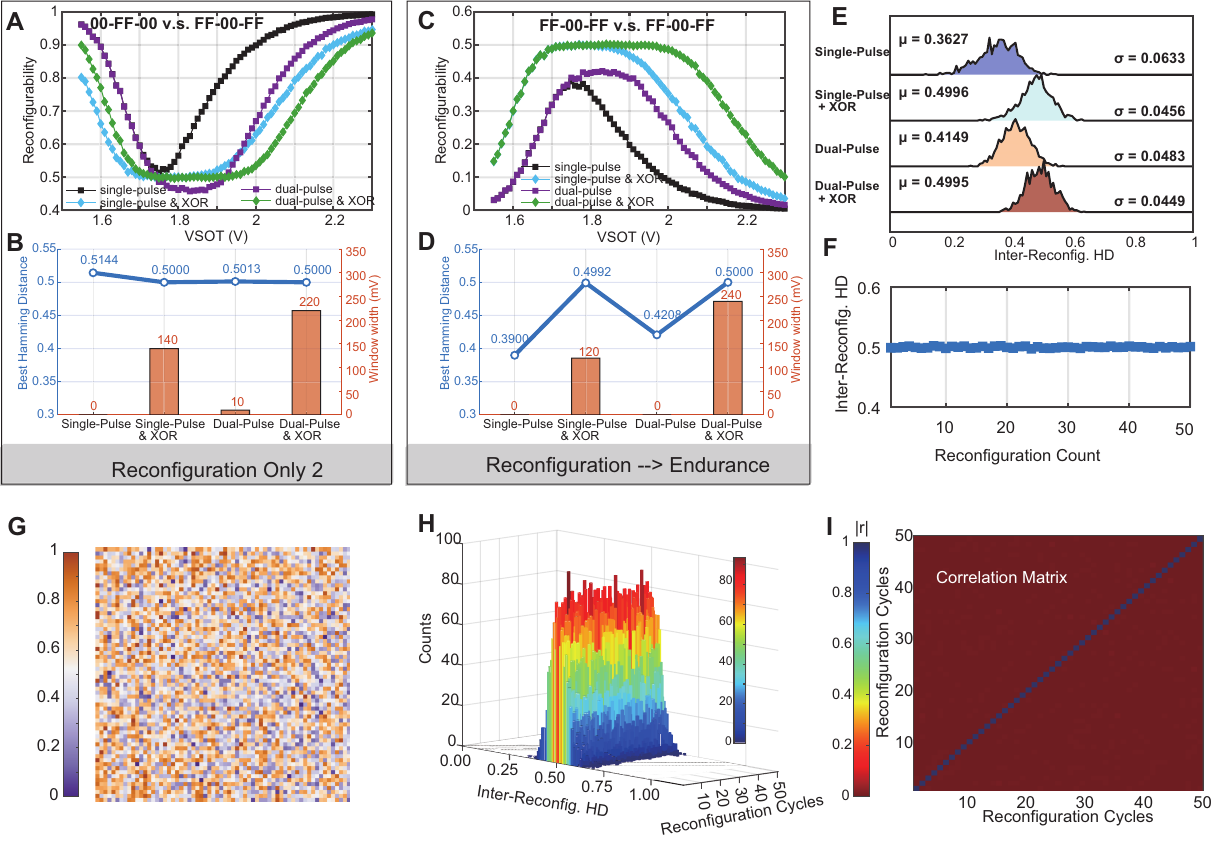}
 \caption{\textbf{Reconfigurability with the proposed dual-pulse reconfiguration strategy.} (\textbf{A-B}) Reconfigurability between 00–FF–00 and FF–00–FF polarized writing. (\textbf{C-D}) Reconfigurability with FF–00–FF writing repeated. (\textbf{E}) Inter-reconfiguration HD using the single- and dual-pulse reconfiguration strategies without/with XOR postprocessing. The write voltage was set to 1.8 V for the single-pulse reconfiguration as well as the first voltage for the dual-pulse reconfiguration. $\beta$ was set to 0.15 V, and the second voltage was set to 1.65 V. (\textbf{F}) Mean value of the inter-reconfiguration HD for 50 reconfigurations. (\textbf{G}) Extracted switching probability map; only 4 kb is shown here. (\textbf{H}) Distribution of the inter-reconfiguration HD for 50 reconfigurations. (\textbf{I}) Correlation matrix of reconfigured keys. The results in subfigures F–I are for FF–00–FF polarized writing.}
 \label{fig:Reconfig}
\end{figure}

The reconfigurability is defined by the inter-reconfiguration HD, which reflects the uniqueness among reconfigured keys. One way to realize reconfigurability is to use the C2C and D2D variability without severe spatial domain dependence. As shown in Figs. \ref{fig:Reconfig}A and \ref{fig:Reconfig}B, 00–FF and FF–00 polarized writing with the single-pulse reconfiguration strategy can achieve two reconfigurations with a near-ideal inter-reconfiguration HD. The dual-pulse reconfiguration strategy can increase the size of the operating window but suffers from the inherent limit that it allows only two reconfigurations, i.e., configured CRP by 00–FF–00 (CRP@00–FF–00) and FF–00–FF (CRP@FF–00–FF) writing. As shown in Figs. \ref{fig:Reconfig}C and \ref{fig:Reconfig}D, the inter-reconfiguration HD between the first-time configuration (CRP@1st-00–FF–00) and the second-time configuration (CRP@2nd-00–FF–00) was 0.4, which is insufficient for the construction of two separate keys. To solve this issue, XOR postprocessing was applied to improve the inter-reconfiguration HD (green lines in Fig. \ref{fig:Reconfig}C). 

The mutual HDs among CRP@1st-00–FF–00, @2nd-00–FF–00, until @N-th-00–FF–00 were sufficient, which indicated that these N sets of CRPs can be considered as independent PUFs. Thus, the proposed dual-pulse reconfiguration strategy achieved nearly unlimited cycling, which leverages the high endurance of SOT-MRAM. Theoretically, $2^{128}$ combinations can be generated in an unpredictable form, and the strong endurance of SOT-MRAM is sufficient to support real-time reconfiguration for the entire life cycle (Fig. \ref{figSI:overall}, Fig. \ref{figSI:PUForPUF}, and Sec. \ref{SecC2CD2D}). Figure \ref{fig:Reconfig}E shows that the reconfigurability with the proposed strategy combining dual-pulse reconfiguration and XOR postprocessing holds a good mean and narrow distribution. Thus, the proposed strategy achieves a large operating window width, ideal inter-reconfiguration HD, and unlimited reconfiguration.

Figure \ref{fig:Reconfig}F shows that the mean value of the inter-reconfiguration HD among 50 reconfigured keys was calculated as 0.5000 within a range of 0.4973$\sim$0.5022, which demonstrates the stability and effectiveness of the proposed strategy. Notably, the strategy supports bidirectional reconfiguration (i.e., multiple writing of FF–00–FF and/or 00–FF–00). The reconfigurability originates from the relatively independent distributions of C2C and D2D variability in the temporal and spatial domains, reinforced by XOR postprocessing, as shown by the switching probability map in Fig. \ref{fig:Reconfig}G. Figure \ref{fig:Reconfig}H presents the distribution of the inter-reconfiguration HD for 50 reconfigured keys. The correlation matrix in Fig. \ref{fig:Reconfig}I confirms the nonlinearity and non-correlation among these reconfigured keys. Generating keys using 00–FF–00 polarized writing also resulted in good reconfigurability and independence (Fig. \ref{figSI:Reconfig0} and \ref{figSI:Reconfig002FF200}). We demonstrate the feasibility and reconfigurability of the proposed rPUF design. The 128 kb capacity of the SOT-MRAM chip enabled a large CRP space and unlimited reconfiguration. The uniqueness, randomness, and reliability of the rPUF were verified at the chip level.


\subsection{Resilience against Temperature}

\begin{figure}[!th]
\centering
 \includegraphics[width=0.7\linewidth]{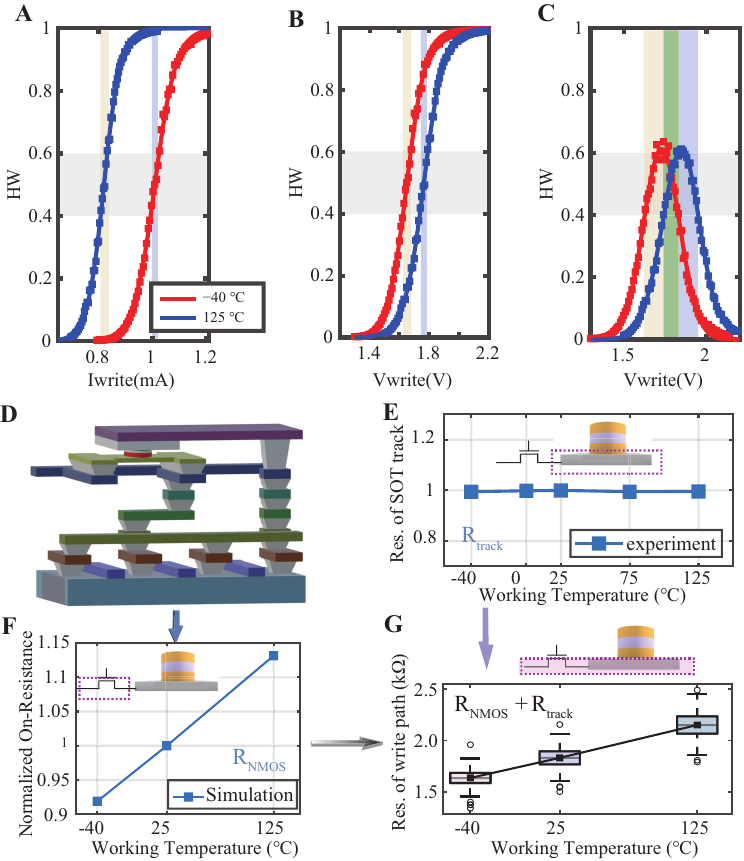}
 \caption{ \textbf{Temperature compensation effect in the rPUF for a SOT-MRAM chip.} Probability curves as functions of the (\textbf{A}) write current and (\textbf{B}) write voltage with the single-pulse reconfiguration strategy and (\textbf{C}) write voltage with the dual-pulse reconfiguration strategy. Experiments were performed at --40°C and 125°C. The write current was calculated from the write voltage and average write path resistance. (\textbf{D}) Three-dimensional layout of the two-transistor one-MTJ (2T1J) structure of the SOT-MRAM chip. (\textbf{E}) SOT track resistance according to the temperature of single MTJs. (\textbf{F}) On-resistance for the write transistor according to temperature. (\textbf{G}) Measured write path resistance.}
 \label{fig:compense}
\end{figure}

Next, the impacts of temperature on the reconfiguration and performance of the proposed rPUF were evaluated. As shown in Fig. \ref{fig:compense}A, the critical switching current is inherently sensitive to the operating temperature and tends to decrease with an increase in temperature. This phenomenon can be attributed to the reduced thermal energy barrier at high temperatures, which facilitates magnetization reversal. As shown in Fig. \ref{fig:compense}B, the single-pulse reconfiguration strategy resulted in smaller deviation in the voltage-dependent curves at --40°C and 125°C, which indicates a smaller difference in the critical write voltage across operating temperatures. However, overlapping operating windows could not be achieved, and uniform pulse settings could not be achieved across the temperature range, even with XOR postprocessing. As shown in Fig. \ref{fig:compense}C, the dual-pulse reconfiguration strategy also resulted in a smaller deviation as well as a common operating window across various temperatures. The second pulse corrected the deviation in the first pulse. Optimizing the combination of the first and second pulses resulted in an operating window that spanned the entire temperature range (green color). Thus, the temperature resilience for the reconfigurability of the proposed rPUF was verified at the chip level.
 
To our knowledge, the temperature compensation effect in CMOS-integrated SOT-MRAM chips has not been explored. Here, we first discuss the prerequisites for the dual-pulse reconfiguration strategy to be feasible and then explain the temperature compensation effect. Based on the range of applicability of XOR postprocessing, the voltage ranges of the target window can be denoted as $\left [a_1, a_2\right ]$ at --40°C and as $\left [b_1, b_2\right ]$ at 125°C. It is a prerequisite that the intersection of these two ranges exists: 
\begin{equation}
\max(a_1, b_1) \leq \min(a_2, b_2) \label{eq:prejudgeCommon}
\end{equation}
In other words, the intersection of the two voltage ranges ($[a_1, a_2] \cap [b_1, b_2]$) must not be empty.

 For the SOT-MRAM chip, a two-transistor-one-MTJ (2T1J) structure was adopted, where each MTJ was associated with a write transistor and read transistor, as shown in Fig. \ref{fig:compense}D \cite{wang2022layout}. The write path resistance mainly comprises the on-resistance of the write transistor and the resistance of the SOT track. Figures \ref{fig:compense}E and \ref{fig:compense}F show the relationships of these two resistances with the temperature. The SOT track resistance was measured for single MTJs fabricated on the same platform. The SOT track resistance of annealed $\beta-$phase tungsten ($\beta-$W) exhibited a weak thermal dependence (Table \ref{TableSI: betaW}), which is consistent with the previous report\cite{hao2015beta}. The transistors exhibited a positive temperature coefficient with the on-resistance increasing with temperature (see Fig. \ref{fig:compense}F). This behavior can be attributed to the reduction in carrier mobility with increasing temperature, which decreased the charge carrier density per unit area in a fixed electric field and thus increased the on-resistance. If the on-resistance and SOT track resistance are combined, then the results indicate that the total write path resistance increases with temperature. As shown in Fig. \ref{fig:compense}G, measurements in analog mode confirmed this positive temperature dependence. Therefore, while the critical write current decreases with temperature, the corresponding deviation in the write voltage is compensated for by an increase in the on-resistance, which stabilizes the voltage-dependent characteristics.

The proposed rPUF addresses both of the main issues of the environmental temperature: real-time reconfiguration and read reliability. The SOT-MRAM chip demonstrated excellent reconfigurability from --40°C to 125°C (Fig. \ref{figSI:Reconfig_Temp}), which proved that the dual-pulse reconfiguration strategy can be used regardless of the operating temperature while increasing the operating window. Thus, the two major bottlenecks concerning real-time reconfiguration are resolved. In terms of read reliability, the SOT-MRAM chip demonstrated improved BER reliability at non-nominal temperatures and power supply VDD conditions (Fig. \ref{figSI-reliability-SWB}), which was enhanced by SWB operation and the differential read circuit compared to the baseline without SWB (Fig. \ref{figSI-reliability}).

\subsection{Numerical modeling}

We developed a numerical model to optimize the pulse combinations for the dual-pulse reconfiguration strategy ($\beta$ = $|V1|$ -- $|V2|$). $\beta$ determines the width of the operating window, which can be defined as the intersection of the WSR and the target window. The final WSR can be expressed as a composite of two WSR functions for the single pulses $WSR1(V)$ and $WSR2(V)$. Because the switching probabilities of two write operations are statistically independent (see Sec. \ref{SecIndep}), the final WSR of the dual-pulse reconfiguration strategy ($F(V1,\beta$)) can be expressed as

\begin{equation}
    F(V1,\beta) = WSR1(V1) - WSR1(V1)\cdot WSR2(V2)
    \label{EqIni}
\end{equation}

\begin{figure}[!h]
 \centering
 \includegraphics[width=0.8\linewidth]{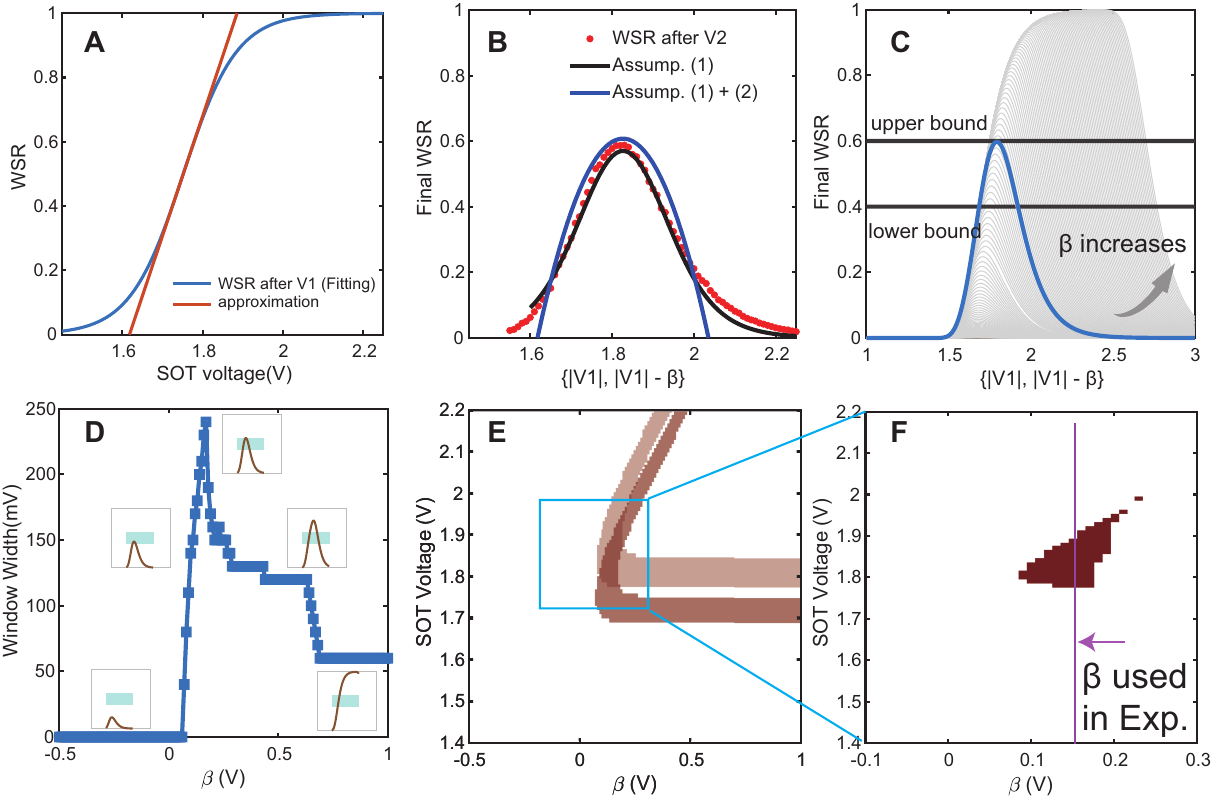}
 \caption{\textbf{Numerical modeling of the dual-pulse reconfiguration strategy and solution for $\beta$.} (\textbf{A}) Simplified model for WSR curves with the single-pulse reconfiguration strategy. (\textbf{B}) Experimental and predicted final WSR with the dual-pulse reconfiguration strategy. (\textbf{C}) Final WSR with the dual-pulse reconfiguration strategy using different values of $\beta$ based on $WSR1$ at room temperature. (\textbf{D}) Width of the operating window as a function of $\beta$. Insets show five typical cases. (\textbf{E}) Phase diagrams of available pulse combinations as functions of $\beta$ and the V1 in the cases of --40°C and 125°C. The overlapped region represents the common operating window across operating temperatures. (\textbf{F}) Magnified view of the common operating window and location of $\beta$ used in experiments (0.15 V).}
 \label{fig5Model}
\end{figure}

Fig. \ref{fig5Model}A shows the WSR function of a single-pulse, which has a sigmoid-like shape and behaves similarly with both 00–FF and FF–00 polarized writing. This results in Assumption 1: Single pulses of different polarities correspond to the same WSR function (i.e., $WSR1(V)\approx WSR2(V)$). Thus, Eq. \ref{EqIni} can be revised as

\begin{equation}
F(V1,\beta) = WSR1(V1) - WSR1(V1)\cdot WSR1(V1 - \beta)
\label{eqAssum1}
\end{equation}
The corresponding result in Fig. \ref{fig5Model}B (black line) shows the close agreement between the numerical model and experimental values, which indicates that Assumption 1 is reasonable. 

The calculation can be further simplified by Assumption 2: Within a specific range, the voltage is linearly related to the WSR with a slope of $k$, which is represented by the tangent line as shown in Fig. \ref{fig5Model}A. $k$ mainly depends on the process variation (see Sec.\ref{SecParameters}). Combining Assumptions 1 and 2 results in
\begin{equation}
F = WSR1(V1) - WSR1(V1)\cdot (WSR1(V1) - k\cdot\beta)
\label{Eq_AB}
\end{equation}
where $WSR1$ reduces to a unitary function. As shown in Fig. \ref{fig5Model}B, the corresponding result (blue line) simplifies to a univariate quadratic curve. In this case, the value of $\beta$ can be determined by solving for the coordinates of the extremes, which results in the optimal value for $\beta$ being 0.147 V (Sec. \ref{SecDualPulse}). Considering the accuracy of the source meter, $\beta$ was set to a value of 0.15 V in the experiment. 

Figure \ref{fig5Model}C presents a parameter sweep of $\beta$ for the dual-pulse reconfiguration strategy, with the model of Eq. \ref{eqAssum1}. Fig. \ref{fig5Model}D summarizes the widths of the operating window with different $\beta$ and the write voltage limited to $<$ 2.4 V. When $\beta$ was below a certain threshold value, no operating window existed. When $\beta$ exceeded the threshold value, the operating window initially increased until it reached a peak width. Further increasing $\beta$ caused the operating window to start decreasing in width. As shown by the inset, the extremes of the $F(V1,\beta)$ curve gradually approached the lower bound of the target window and then the upper bound before overflowing the target window. 
 
 Based on the temperature range, phase diagrams of available pulse combinations as a function of $\beta$ versus $V1$ were generated for --40°C and 125°C and overlapped, as shown in Fig. \ref{fig5Model}E. Colored pixels indicate where $F(V1,\beta)$ lies within the target window, while the overlapping region represents the common operating window across temperatures. Figure \ref{fig5Model}F shows a magnified view of the common operating window. Both the calculated value (0.147 V) and experimental value (0.15 V) of $\beta$ are within this region and close to the central area, which demonstrates the strong agreement between the experimental data and simulation results. In addition, the experimental results also showed that the $\beta$ is consistent and universal for the same batch of chips (Table \ref{TableSI:10chips} and Sec. \ref{SecParameters}), which means that customized designs for each chip can be avoided. With the prospect of fully functional chips (Fig. \ref{figSI:structure}), the dual-pulse reconfiguration strategy can be implemented by the bandgap reference and reference buffer circuits, which are both industry-standard circuit modules.

\section{Discussion}

We developed an rPUF for CMOS-integrated SOT-MRAM chips that can operate over a wide temperature range and reconfigure in real-time, which is a pivotal advance for hardware-based security primitives. The proposed dual-pulse reconfiguration strategy modifies the monotonically increasing relationship between the switching probability and electrical excitation of the conventional single-pulse strategy to increase the size of the operating window. The CMOS-integrated SOT-MRAM chip offers a unique temperature compensation effect that facilitates real-time reconfiguration from --40°C to 125°C. With increasing temperature, the increased write path resistance counteracts the decreased critical write current of the MTJ, which results in a smaller offset of the critical write voltage compared to that of the critical write current. The proposed design demonstrated notable improvements in various performance metrics, which highlights its potential application to improving the reliability of hardware-based security. The C2C and D2D variability was comprehensively investigated, and the results showed that combining the dual-pulse reconfiguration strategy with XOR postprocessing resulted in robustness against machine learning attacks and excellent reconfigurability. Meanwhile, the high-level endurance of the SOT-MRAM chip supports high-order reconfiguration counts. The read reliability is enhanced by a differential read circuit and SWB, which results in an approximately infinite inter/intra-HD ratio and ultralow BER. As a forward-looking solution, our rPUF provides scalable reconfigurability in both the spatial domain (from 1 b to 128 kb) and the temporal domain (on-demand refresh of CRPs) and thus provides a flexible security solution for dynamic operating environments. The design mitigates temperature-induced write reconfiguration issues while providing a cost-effective solution for hardware-based security and thus paves the way for implementation in IoT applications.

\section{Materials and methods \label{SecEXP}}

\subsection*{Fabrication Processes}

The film stack and SOT-MRAM chip were fabricated in-house. The optimized W/CoFeB/MgO/CoFeB/SAF system was adopted for the film stack \cite{jiang2023demonstration}. A top-pinned structure was adopted for the MTJ, which was prepared above M5V5 as the last level of metal and via in the front-end-of-line (FEOL) substrate. The pillars of the MTJ had a short axis of $\sim$300 nm and an aspect ratio of $\sim$2. One basic unit comprised two MTJs for storage and reading. The film stack could tolerate temperatures up to 400°C, which is typical for solder reflow. The SOT track passed the electro-migration (EM) and stress migration tests \cite{xu2023full}. The chemico-mechanical polishing process was used to reduce the roughness of the FEOL substrate for deposition of a high-quality film stack \cite{zhang2022integration}. The etching process was improved by adjusting the plasma conditions and etching angle to alleviate the footing issue. A fine device morphology was achieved, and redeposition of the etch on the sidewall was avoided. At the chip level, the retention was $>$ 10 years. The wafer-level endurance was more than $1\times10^{14}$ \cite{xu2023full}.

\subsection*{Electrical Measurement} 

The 128 kb SOT-MRAM chip supported both digital and analog signal modes. The resistances of the read and write paths for each address were tested in analog signal mode. Under normalized operation conditions, the current flowing through the selected basic unit was calculated and then converted to the resistance of the selected path. The overall array was characterized, and the statistical values were finally obtained. The different write voltages were applied via the analog I/O. The data were read in digital signal mode. Both modes supported variable temperature measurements. All evaluations at the chip level were carried out by using the chip-probing test platform.

\section*{Supplementary material} 

Supplementary material for this article is available online.

\bibliography{reference} 
\bibliographystyle{sciencemag}

\section*{Acknowledgments} 
\textbf{Funding:} This work was supported by the National Key Research and Development Program of China (Grant Nos. 2021YFB3601303) and National Natural Science Foundation of China (No. 62171013).

\textbf{Author contributions:} M. W., C. J., and Z. W. conceived and designed the experiments. C. J., H. L., and Z. Z., conducted the experiment and data acquisition. M. W., W. Z., Z. H., and Y. Z. analyzed the data. All authors discussed the results. M. W. wrote the draft, and all authors contributed to the review and editing. Z. W., H. L. and W. Z. supervised the project. 

\textbf{Competing interests:} The authors declare no conflict of interest. 

\textbf{Data Availability Statement:} All data are available in the main text or the supplementary materials.

\clearpage

\end{document}


\title{\centering Supporting Information \\ ~ \\ Reconfigurable Physical Unclonable Function based on SOT-MRAM Chips}
	
	%
	
	\maketitle
	
	
	
	\clearpage
	
	\setcounter{figure}{0}
	\setcounter{table}{0}
	\setcounter{equation}{0}
	\setcounter{section}{0}
	
	\renewcommand{\thefigure}{S\arabic{figure}}
	\renewcommand{\thetable}{S\arabic{table}}
	\renewcommand{\thesection}{S\arabic{section}}
	\renewcommand{\theequation}{S\arabic{equation}}

	\tableofcontents
	
	\clearpage
	
	\section{Write shmoo of the SOT-MRAM chip}
	
\begin{figure}[!h]
		\centering
		\includegraphics[width=1\linewidth]{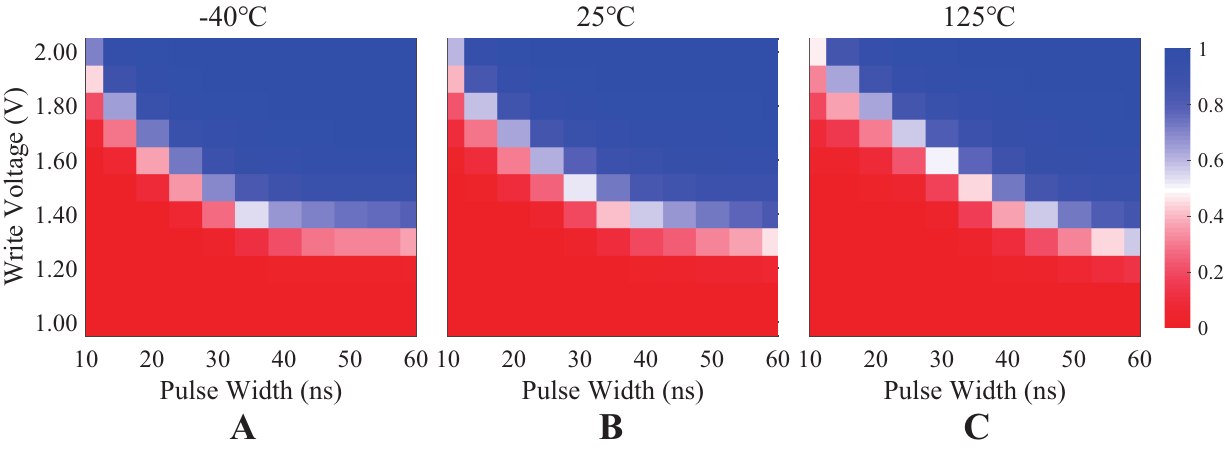}
		\caption{\textbf{Write Shmoo of SOT-MRAM chip under various pulse widths and write voltages.} Working temperature: (\textbf{A}) -40 ℃; (\textbf{B}) 25 ℃; (\textbf{C}) 125 ℃. The color-mapping refers to the write success rate (WSR).}
		\label{figSI:Shmoo}
	\end{figure}

	We conducted writing/reading operations under various working temperatures, ranging from -40 ℃ to 125 ℃, which aligns with the criteria for industrial products. Figure \ref{figSI:Shmoo} presents the experimental results of the SOT-MRAM chip in multiplexer testing mode, demonstrating the chip’s functionality across the temperature range.
	
	\clearpage
	\section{Introduction of XOR operation\label{secXOR}}
	
	XOR-based post-processing is a widely used technique for improving PUF metrics, such as uniformity, uniqueness, etc. XOR operations with different bit numbers influence uniformity evaluation, where $\mu$ and $\sigma$ are obtained by fitting the histogram. A response is defined as 128 bit, thus the ideal $\mu$ and $\sigma$ are:
	\begin{equation}
	Ideal ~\mu = 0.5
	\end{equation}
	
	\begin{equation}
	Ideal~ \sigma = \frac{1}{2\sqrt{128}} \simeq 0.0442
	\end{equation}
	
\begin{figure}[!h]
		\centering
		\includegraphics[width=0.7\linewidth]{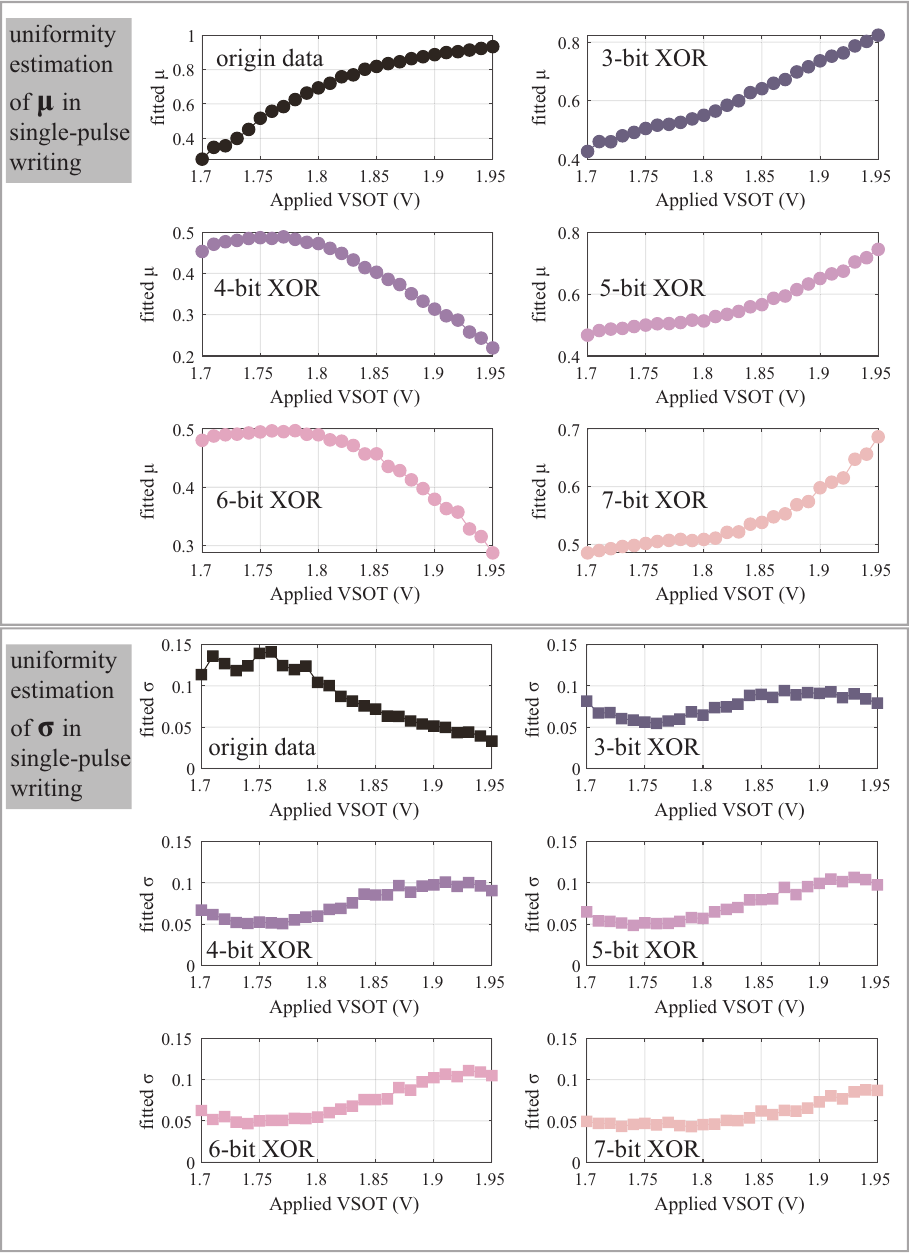}
		\caption{Fitted $\mu$ and $\sigma$ in uniformity evaluation with various XOR processing. The single-pulse writing strategy is applied.}
		\label{figSI:XORmethod_V1}
	\end{figure}

\begin{figure}[!h]
		\centering
		\includegraphics[width=0.7\linewidth]{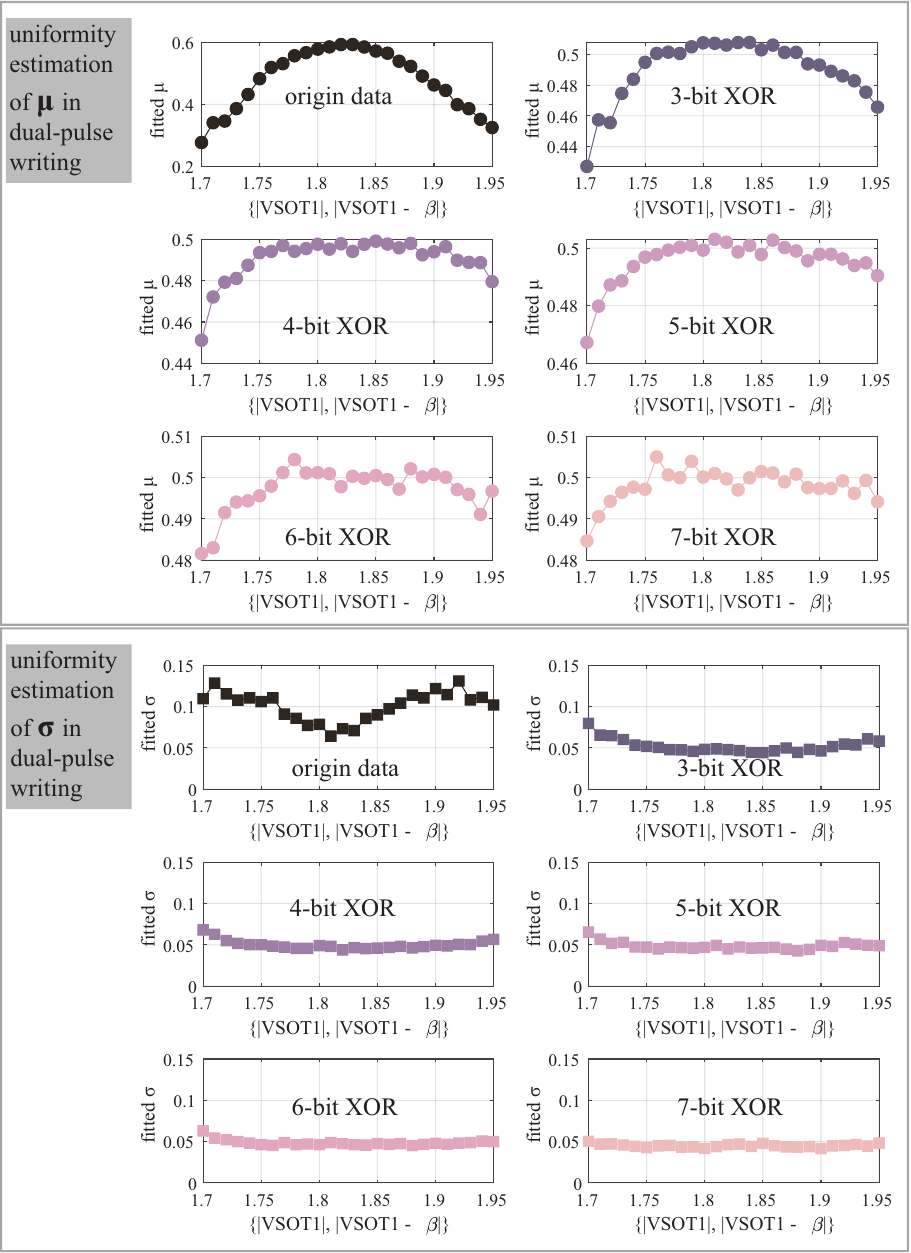}
		\caption{Fitted $\mu$ and $\sigma$ in uniformity evaluation with various XOR processing. The dual-pulse writing strategy is applied.}
		\label{figSI:XORmethod_V2}
	\end{figure}

	Figure \ref{figSI:XORmethod_V1} and \ref{figSI:XORmethod_V2} refer to uniformity evaluation of PUF responses generated by the single-pulse writing strategy and dual-pulse writing strategy, respectively. It could be observed that:
	
	Firstly, as the bit number in the XOR operation increases, both $\sigma$ and $\mu$ show a trend of improvement. Meanwhile, it should be noted that the modified results also show a slight fluctuation according to the original data. 
	
	Secondly, the even/odd dependence. When the number of bits is even, the XOR operation eliminates the counting of parallel and anti-parallel states. For example, $1\oplus 1 = 0$, $0\oplus 0 = 0$. Conversely, when the number of bits is odd, this difference is preserved. For instance, $1\oplus 1\oplus 1 = 1$, $0\oplus 0\oplus 0 = 0$. In wide-temperature-range measurement, the 7-bit XOR operation is finally adopted.
	
	Thirdly, the uniformity evaluation in the case of dual-pulse writing (see Fig. \ref{figSI:XORmethod_V2}) shows better performance compared with that in single-pulse writing (see Fig. \ref{figSI:XORmethod_V1}).
	
	\clearpage
	\section{Robustness against machine learning attacks \label{SecMLattacks}}
	
	Five ML attacks are investigated, including support vector machine (SVM), random forest, artificial neural network (ANN), logistic regression (LR), and Covariance Matrix Adaptation Evolution Strategy (CMA-ES). In terms of machine learning attacks, the PUF data is randomly divided into a training set and a test set, and the ratio of the two is 7:3. The training set is used for training and iteration, and the test set is used to determine the accuracy of machine learning predictions. The following are the details of those attack methods.
	
	\begin{figure}[!h]
		\centering
		\includegraphics[width=0.7\linewidth]{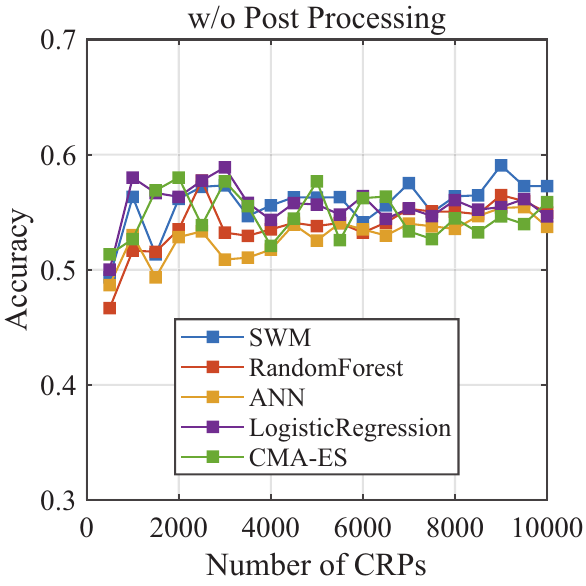}
		\caption{Machine learning results for the training data without XOR operation.}
		\label{figSI:ML}
	\end{figure}
	
	1. Logic Regression (LR) Attack. The LR attack uses a sigmoid function to discriminate the output and improve the accuracy of the prediction. Ridge regression is used to prevent overfitting, and the LBFGS algorithm optimizes the parameters used in the model.
	
	2. Support Vector Machine (SVM) Attack. SVM achieves partitioning of data by finding an optimal hyperplane. The kernel function is set to Radial Basis Function (RBF). Regularization parameter and Kernel coefficient are determined through grid search with 3-fold cross-validation.
	
	3. Covariance Matrix Adaptation Evolution Strategy (CAM-ES). During the iterative process, suitable parameters and models will “naturally evolve”. The core approach is to evolve the model towards an optimal solution through parameter tuning. The number of XOR operations used in each stage is 1, which is the typical operation in the attack against PUF. Fitness function refers to the method introduced in Ref. \cite{xu2023modeling}.
	
	4. Artificial Neural Network (ANN) Attack. ANN mimics the connections between biological neurons. Here we employ the Multi-Layer Perceptron (MLP) classifier, consisting of the input layer, the hidden layer with 100 neurons, and the output layer. This method uses L2 regularization, ReLu activation function, and LBFGS as the solver.
	
	5. Random Forest (RF) Attack. Based on the Bagging algorithm, RF frequently creates multiple new classifiers, forming decision trees. The measure of split quality uses information gain (entropy). The out-of-bag (OOB) samples are allowed for assessment of model performance.
	
	The ML attack results for data without XOR post-processing are evaluated as 0.5529 (see Fig. \ref{figSI:ML}).

	\clearpage
	\section{Main metrics of PUF design\label{SecUniformity}}
	
	The following presents the evaluation of the uniformity of the SOT-MRAM rPUF, with a PUF response defined as 128 bits. 
	
\begin{figure}[!h]
		\centering
		\includegraphics[width=0.8\linewidth]{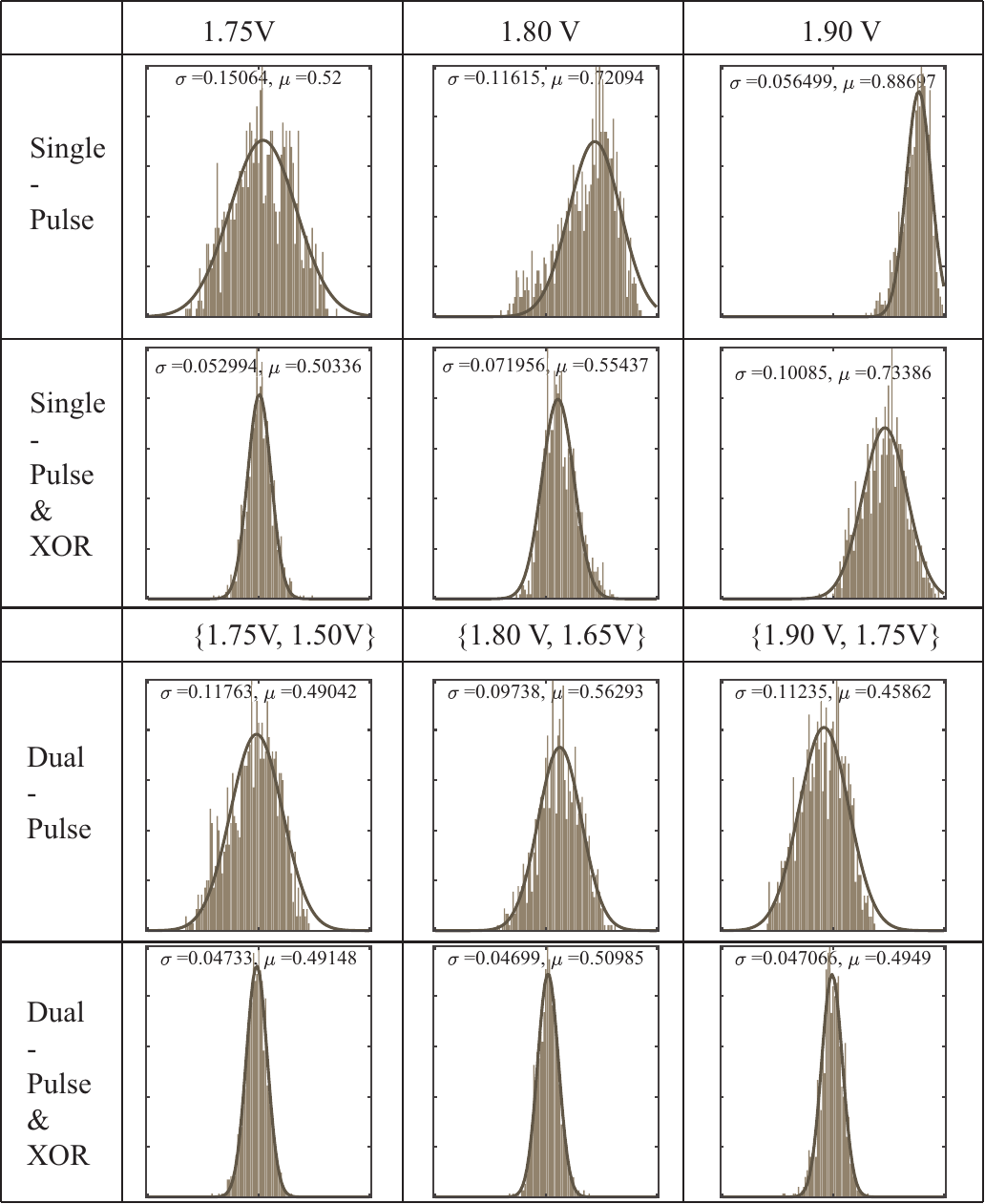}
		\caption{Uniformity of SOT-MRAM rPUF in 00-FF and 00-FF-00 polarized reconfiguration.}
		\label{figSI:Uniformity_002FF200}
	\end{figure}

	Figure \ref{figSI:Uniformity_002FF200} refers to the uniformity evaluation of 00-FF polarized single-pulse and 00-FF-00 polarized dual-pulse reconfiguration. 
	
\begin{figure}[!h]
		\centering
		\includegraphics[width=0.8\linewidth]{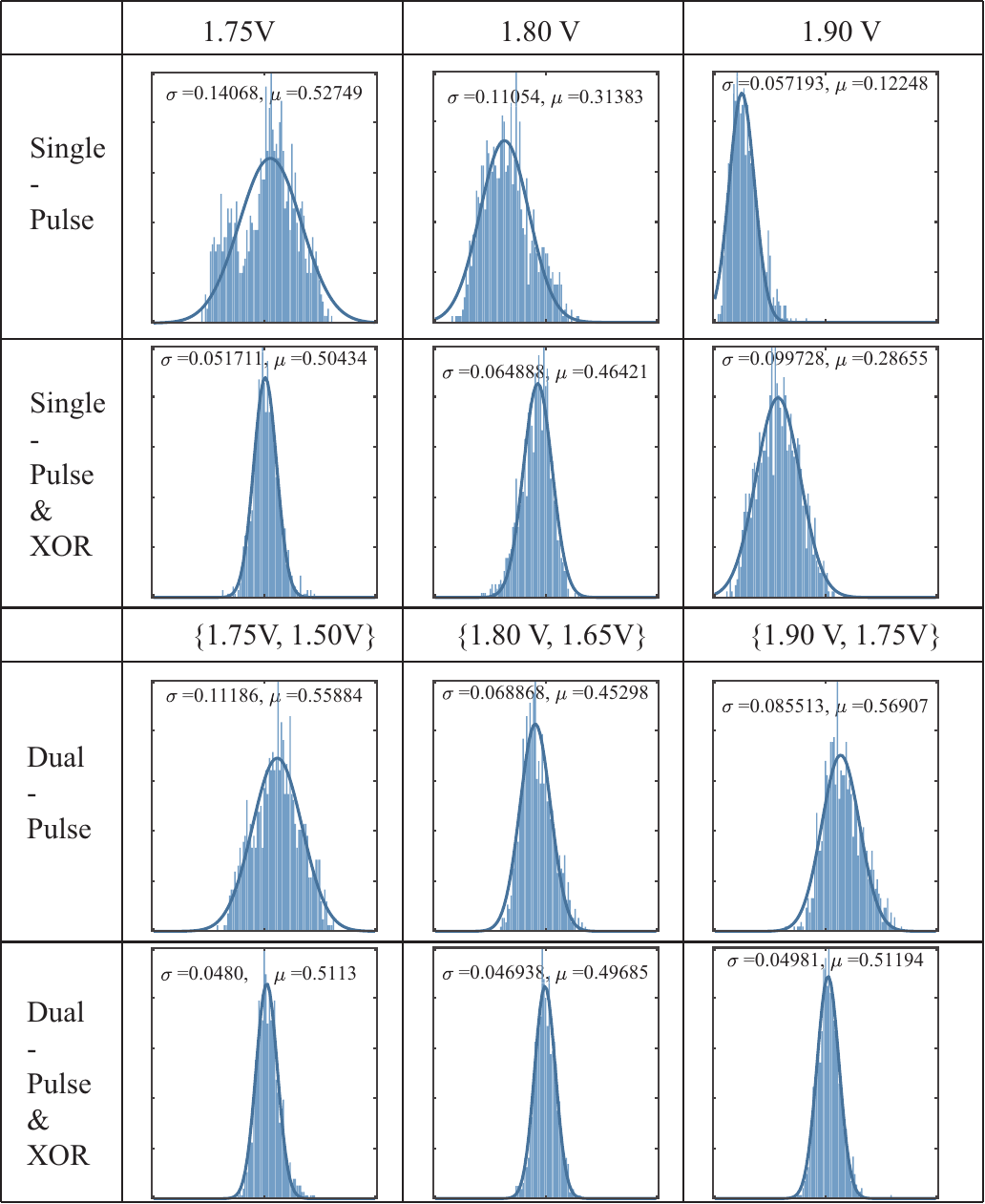}
		\caption{Uniformity of SOT-MRAM rPUF in FF-00 and FF-00-FF polarized reconfiguration.}
		\label{figSI:Uniformity_FF2002FF}
	\end{figure}

	Figure \ref{figSI:Uniformity_FF2002FF} refers to the uniformity evaluation after FF-00 polarized single-pulse and FF-00-FF polarized dual-pulse reconfiguration. 
	
	It is observed that the dual-pulse writing effectively broadened the operation window width. The uniformity of PUF responses is further enhanced via the XOR operation. As a result, more intensive distributions and reduced  $\sigma$ in Gauss fitting are achieved by the dual-pulse method and XOR operation in Fig. \ref{figSI:Uniformity_002FF200} and \ref{figSI:Uniformity_FF2002FF}. Here, the 3-bit XOR operation is used to show a clear trend in $sigma$ values. In the use of 7-bit XOR, the uniformity after dual-pulse writing is estimated as 0.5001 as fitted $\mu$, with 0.042 as fitted $\sigma$.

	\clearpage
	
\begin{figure}[!h]
		\centering
		\includegraphics[width=0.8\linewidth]{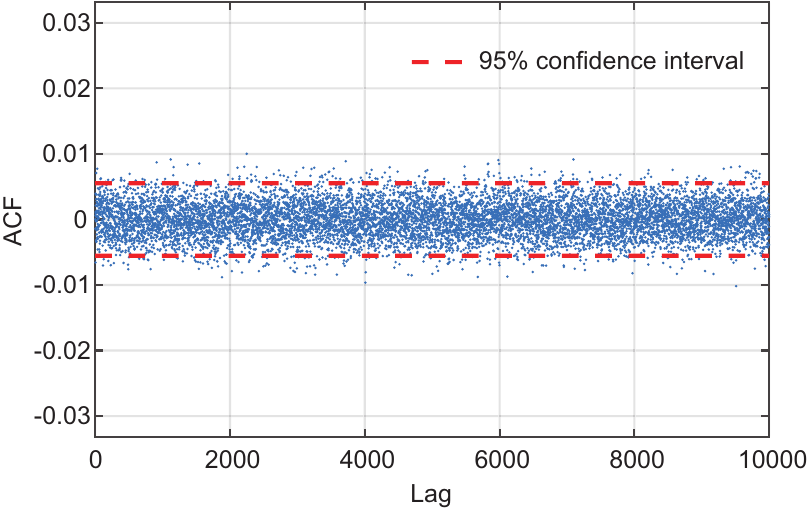}
		\caption{Auto-correlation Function (ACF) testing of the measured PUF responses.}
		\label{figSI:ACF}
	\end{figure}

	Fig. \ref{figSI:ACF} shows the result of the Auto-Correction Function (ACF) test, where the majority of data points are located within the confidence range interval. The ACF test result confirms the unpredictability and randomness of reconfiguration in our proposed rPUF design.
	
	Table \ref{TabSI:NIST} lists the NIST SP800-22 test result, confirming the randomness of reconfigured CRPs by the proposed rPUF design.
	\begin{table}
		\centering
		\caption{NIST SP800-22 test result}
		\label{TabSI:NIST}
		\begin{tabular}[htbp]{@{}llll@{}}
			\hline
			Test                & p-Value  & Proportion & Pass? \\
			\hline
			Frequency           & 0.122325 & 10/10      & YES  \\
			BlockFrequency      & 0.911413 & 10/10      & YES  \\
			CumulativeSums-1    & 0.911413 & 9/10       & YES  \\
			CumulativeSums-2    & 0.911413 & 9/10       & YES  \\
			Runs                & 0.534146 & 9/10       & YES  \\
			LongestRun          & 0.122325 & 10/10      & YES  \\
			Rank                & 0.350485 & 10/10      & YES  \\
			FFT                 & 0.534146 & 10/10      & YES  \\
			NonOverlappingTemplate & PASS  & PASS       & YES  \\
			OverlappingTemplate & 0.350485 & 9/10       & YES  \\
			ApproximateEntropy  & 0.017912 & 10/10      & YES \\
			Serial-1            & 0.534146 & 10/10      & YES  \\
			Serial-2            & 0.534146 & 10/10      & YES  \\
			LinearComplexity    & 0.350485 &  9/10      & YES \\
			\hline
		\end{tabular}
		\label{Table:NIST}
	\end{table}
	
\begin{figure}[!h]
		\centering
		\includegraphics[width=0.8\linewidth]{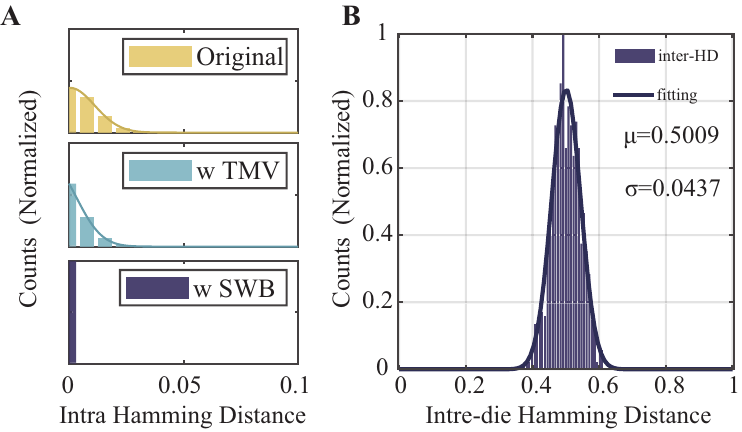}
		\caption{(\textbf{A}) Intra HD: direct readout, time-majority-vote (TMV), and self-write-back (SWB). (\textbf{B}) Inter-die HD distribution with the mean value near-ideal 50\%.}
		\label{figSI:Intra_Inter}
	\end{figure}

	The intra-HD and inter-die HD are shown in Fig. \ref{figSI:Intra_Inter}. Fig. \ref{figSI:Intra_Inter}A shows 3 cases of intra Hamming distance (HD): direct readout, time-majority-vote (TMV), and self-write-back (SWB). Fig. \ref{figSI:Intra_Inter}B exhibits good uniqueness, evaluated by the calculation of the inter-die HD. The Inter/Intra-HD ratio approximates $\infty$ with the aid of the SWB technique, estimated as $\sim$1500 $\times$.
	
	\clearpage
	\section{Illustration of the origin of reconfigurability}
	
	\subsection{Comprehension of C2C and D2D variability}
	The spatial distribution of PUF data is the union of all units with certain time-domain data. In the performance of reconfiguration, device-to-device (D2D) variation shows a mapping relationship with the statistical behavior of cycle-to-cycle (C2C) variation. 
	
\begin{figure}[!h]
		\centering
		\includegraphics[width=0.75\linewidth]{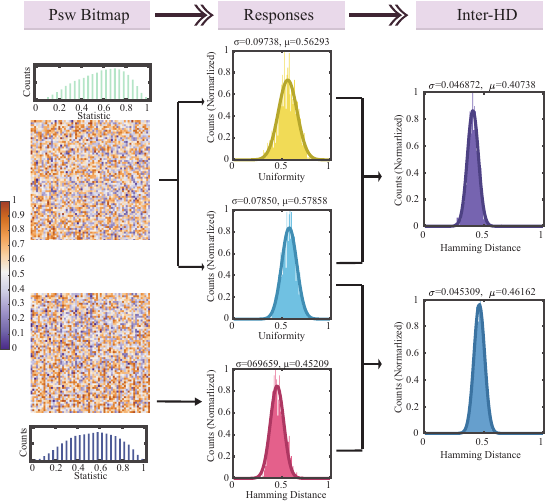}
		\caption{Relationship between Psw statistic map, uniformity of responses, and Inter-Reconfig. HD.} 
		\label{figSI:overall}
	\end{figure}

	In Fig. \ref{figSI:overall}, the first column displays two statistical maps of switching probabilities, with polarity of 00-FF-00 and FF-00-FF, respectively. The switching probabilities for each unit are derived from 50 reconfigurations. At one of these reconfigurations, the chip presents a binary PUF data distribution. The second column counts the uniformity of the PUF response after a certain reconfiguration. The first two rows of the second column are from the raw data based on the first statistical map. The third row of the second column corresponds to the second statistical map. The third column indicates the uniqueness between reconfigurations guided by the connected arrow.

	\begin{figure}[!h]
		\centering
		\includegraphics[width=0.45\linewidth]{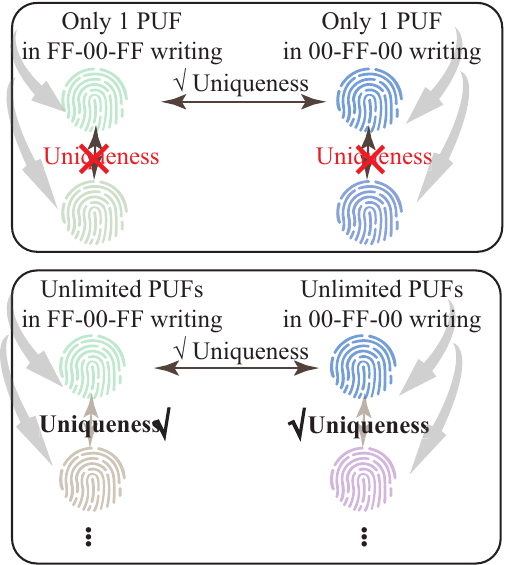}
		\caption{Unlimited PUF reconfiguration in aid of XOR operation.}
		\label{figSI:PUForPUF}
	\end{figure}
	
	The reconfiguration originating from different Psw map results shows better Inter-Reconfig. HD. The Inter-Reconfig. HD. by the same-polarized writing (i.e., from the same Psw map) is not exactly ideal; meanwhile, this result provides a prerequisite for XOR operations. Under the aid of the XOR operation, reconfigurations originating from different/same Psw maps are both available for the formation of separate PUFs. As shown in Fig. \ref{figSI:PUForPUF}, the SOT-MRAM rPUF was successfully validated for unlimited reconfiguration.
	
	The above descriptions and illustrations help the understanding of Fig. 3 in the main manuscript.
	
	\clearpage
	\subsection{Quantitative evaluation of reconfigurability \label{SecC2CD2D}}
	Table \ref{TabSI:Same_WO} refers to the evaluation of reconfigurability: two cases of PUF responses generated by the same polarity but without XOR.
	
	Table \ref{TabSI:Same_W} refers to the evaluation of reconfigurability: two cases of PUF responses generated by the same polarity and with XOR.
	
	Table \ref{TabSI:Diff_WO} refers to the evaluation of reconfigurability: two cases of PUF responses generated by the opposite polarity but without XOR.
	
	Table \ref{TabSI:Diff_W} refers to the evaluation of reconfigurability: two cases of PUF responses generated by the opposite polarity and with XOR.
	
	\begin{table}[h]
		\centering
		\caption{Reconfigurations with the same polarity but without XOR}
		\label{TabSI:Same_WO}
		\renewcommand{\arraystretch}{1.5} 
		\begin{tabular}{ccccccc}
			\toprule
			\multicolumn{1}{c}{Method} & \multicolumn{2}{c}{\makecell[t]{Statistic of Psw \\@case1}} & \multicolumn{2}{c}{\makecell[t]{Statistic of Psw \\@case2}}  & \multicolumn{1}{c}{\makecell[t]{Inter-reconfig.  \\HD }} & \multicolumn{1}{c}{\makecell[t]{Distance \\ from 0.5}} \\ \midrule
			\multicolumn{1}{c}{\multirow{2}{*}{Single-Pulse}} & \multicolumn{2}{c}{00-FF} & \multicolumn{2}{c}{00-FF} & \multicolumn{1}{c}{\multirow{2}{*}{0.3340}} & \multicolumn{1}{c}{\multirow{2}{*}{0.166}} \\
			\multicolumn{1}{c}{} & \multicolumn{1}{c}{$\mu$ = 0.7459} & \multicolumn{1}{c}{$\sigma$=0.2048} & \multicolumn{1}{c}{$\mu$ = 0.7459} & \multicolumn{1}{c}{$\sigma$=0.2048} & \multicolumn{1}{c}{} & \multicolumn{1}{c}{} \\ 
			\multicolumn{1}{c}{\multirow{2}{*}{Dual-Pulse}} & \multicolumn{2}{c}{00-FF-00} & \multicolumn{2}{c}{00-FF-00} & \multicolumn{1}{c}{\multirow{2}{*}{0.4076}} & \multicolumn{1}{c}{\multirow{2}{*}{0.0924}} \\ 
			\multicolumn{1}{c}{} & \multicolumn{1}{c}{$\mu$ = 0.5734} & \multicolumn{1}{c}{$\sigma$=0.2057} & \multicolumn{1}{c}{$\mu$ = 0.5734} & \multicolumn{1}{c}{$\sigma$=0.2057} & \multicolumn{1}{c}{} & \multicolumn{1}{c}{} \\ \hline
			\multicolumn{1}{l}{} &  &  &  &  & \multicolumn{1}{l}{} & \multicolumn{1}{l}{} \\ \hline
			\multicolumn{1}{c}{\multirow{2}{*}{Single-Pulse}} & \multicolumn{2}{c}{FF-00} & \multicolumn{2}{c}{FF-00} & \multicolumn{1}{c}{\multirow{2}{*}{0.3556}} & \multicolumn{1}{c}{\multirow{2}{*}{0.1444}} \\ 
			\multicolumn{1}{c}{} & \multicolumn{1}{c}{$\mu$ = 0.7183} & \multicolumn{1}{c}{$\sigma$=0.2154} & \multicolumn{1}{c}{$\mu$ = 0.7183} & \multicolumn{1}{c}{$\sigma$=0.2154} & \multicolumn{1}{c}{} & \multicolumn{1}{c}{} \\ 
			\multicolumn{1}{c}{\multirow{2}{*}{Dual-Pulse}} & \multicolumn{2}{c}{FF-00-FF} & \multicolumn{2}{c}{FF-00-FF} & \multicolumn{1}{c}{\multirow{2}{*}{0.4152}} & \multicolumn{1}{c}{\multirow{2}{*}{0.0848}} \\ 
			\multicolumn{1}{c}{} & \multicolumn{1}{c}{$\mu$ = 0.5356} & \multicolumn{1}{c}{$\sigma$=0.2120} & \multicolumn{1}{c}{$\mu$ = 0.5356} & \multicolumn{1}{c}{$\sigma$=0.2120} & \multicolumn{1}{c}{} & \multicolumn{1}{c}{} \\ \bottomrule
		\end{tabular}%
	\end{table}
	
	
	\clearpage
	
	\begin{table}[!h]
		\centering
		\caption{Reconfigurations with the same polarity and with XOR}
		\label{TabSI:Same_W}
		\renewcommand{\arraystretch}{1.5} 
		\begin{tabular}{ccccccc}
			\toprule
			\multicolumn{1}{c}{Method} & \multicolumn{2}{c}{\makecell[t]{Statistic of Psw \\@case1}} & \multicolumn{2}{c}{\makecell[t]{Statistic of Psw \\@case2}}  & \multicolumn{1}{c}{\makecell[t]{Inter-reconfig.  \\HD}} & \multicolumn{1}{c}{\makecell[t]{Distance \\ from 0.5}} \\ \midrule
			\multicolumn{1}{c}{\multirow{2}{*}{\makecell[t]{Single-Pulse \\ + XOR}}} & \multicolumn{2}{c}{00-FF} & \multicolumn{2}{c}{FF-00} & \multicolumn{1}{c}{\multirow{2}{*}{0.4709}} & \multicolumn{1}{c}{\multirow{2}{*}{0.0291}} \\
			\multicolumn{1}{c}{} & \multicolumn{1}{c}{$\mu$ = 0.5987} & \multicolumn{1}{c}{$\sigma$=0.1405} & \multicolumn{1}{c}{$\mu$ = 0.5987} & \multicolumn{1}{c}{$\sigma$=0.1405} & \multicolumn{1}{c}{} & \multicolumn{1}{c}{} \\ 
			\multicolumn{1}{c}{\multirow{2}{*}{\makecell[t]{Dual-Pulse \\ + XOR}}} & \multicolumn{2}{c}{00-FF-00} & \multicolumn{2}{c}{FF-00-FF} & \multicolumn{1}{c}{\multirow{2}{*}{0.4972}} & \multicolumn{1}{c}{\multirow{2}{*}{0.0028}} \\ 
			\multicolumn{1}{c}{} & \multicolumn{1}{c}{$\mu$ = 0.5094} & \multicolumn{1}{c}{$\sigma$=0.0818} & \multicolumn{1}{c}{$\mu$ = 0.5094} & \multicolumn{1}{c}{$\sigma$=0.0818} & \multicolumn{1}{c}{} & \multicolumn{1}{c}{} \\ \hline
			\multicolumn{1}{l}{} &  &  &  &  & \multicolumn{1}{l}{} & \multicolumn{1}{l}{} \\ \hline
			\multicolumn{1}{c}{\multirow{2}{*}{\makecell[t]{Single-Pulse \\ + XOR}}} & \multicolumn{2}{c}{FF-00} & \multicolumn{2}{c}{FF-00} & \multicolumn{1}{c}{\multirow{2}{*}{0.4813}} & \multicolumn{1}{c}{\multirow{2}{*}{0.0187}} \\ 
			\multicolumn{1}{c}{} & \multicolumn{1}{c}{$\mu$ = 0.5772} & \multicolumn{1}{c}{$\sigma$=0.1319} & \multicolumn{1}{c}{$\mu$ = 0.5772} & \multicolumn{1}{c}{$\sigma$=0.1319} & \multicolumn{1}{c}{} & \multicolumn{1}{c}{} \\ 
			\multicolumn{1}{c}{\multirow{2}{*}{\makecell[t]{Dual-Pulse \\ + XOR}}} & \multicolumn{2}{c}{FF-00-FF} & \multicolumn{2}{c}{FF-00-FF} & \multicolumn{1}{c}{\multirow{2}{*}{0.4957}} & \multicolumn{1}{c}{\multirow{2}{*}{0.0043}} \\ 
			\multicolumn{1}{c}{} & \multicolumn{1}{c}{$\mu$ = 0.5040} & \multicolumn{1}{c}{$\sigma$=0.0819} & \multicolumn{1}{c}{$\mu$ = 0.5040} & \multicolumn{1}{c}{$\sigma$=0.0819} & \multicolumn{1}{c}{} & \multicolumn{1}{c}{} \\ \bottomrule
		\end{tabular}%
	\end{table}

	\clearpage
	\begin{table}[!h]
		\centering
		\caption{Reconfigurations with the opposite polarity but without XOR}
		\label{TabSI:Diff_WO}
		\renewcommand{\arraystretch}{1.5} 
		\begin{tabular}{ccccccc}
			\toprule
			\multicolumn{1}{c}{Method} & \multicolumn{2}{c}{\makecell[t]{Statistic of Psw \\@case1}} & \multicolumn{2}{c}{\makecell[t]{Statistic of Psw \\@case2}}  & \multicolumn{1}{c}{\makecell[t]{Inter-reconfig.  \\HD}} & \multicolumn{1}{c}{\makecell[t]{Distance \\ from 0.5}} \\ \midrule
			\multicolumn{1}{c}{\multirow{2}{*}{Single-Pulse}} & \multicolumn{2}{c}{00-FF} & \multicolumn{2}{c}{00-FF} & \multicolumn{1}{c}{\multirow{2}{*}{0.5657}} & \multicolumn{1}{c}{\multirow{2}{*}{0.0657}} \\
			\multicolumn{1}{c}{} & \multicolumn{1}{c}{$\mu$ = 0.7459} & \multicolumn{1}{c}{$\sigma$=0.2048} & \multicolumn{1}{c}{$\mu$ = 0.7183} & \multicolumn{1}{c}{$\sigma$=0.2154} & \multicolumn{1}{c}{} & \multicolumn{1}{c}{} \\ 
			\multicolumn{1}{c}{\multirow{2}{*}{Dual-Pulse}} & \multicolumn{2}{c}{00-FF-00} & \multicolumn{2}{c}{00-FF-00} & \multicolumn{1}{c}{\multirow{2}{*}{0.4619}} & \multicolumn{1}{c}{\multirow{2}{*}{0.0381}} \\ 
			\multicolumn{1}{c}{} & \multicolumn{1}{c}{$\mu$ = 0.5734} & \multicolumn{1}{c}{$\sigma$=0.2057} & \multicolumn{1}{c}{$\mu$ = 0.5346} & \multicolumn{1}{c}{$\sigma$=0.2120} & \multicolumn{1}{c}{} & \multicolumn{1}{c}{} \\ \bottomrule
		\end{tabular}%
	\end{table}
	
	\vspace{3cm}
	
	\begin{table}[!h]
		\centering
		\caption{Reconfigurations with the opposite polarity and with XOR}
		\label{TabSI:Diff_W}
		\renewcommand{\arraystretch}{1.5} 
		\begin{tabular}{ccccccc}
			\toprule
			\multicolumn{1}{c}{Method} & \multicolumn{2}{c}{\makecell[t]{Statistic of Psw \\@case1}} & \multicolumn{2}{c}{\makecell[t]{Statistic of Psw \\@case2}}  & \multicolumn{1}{c}{\makecell[t]{Inter-reconfig.  \\HD}} & \multicolumn{1}{c}{\makecell[t]{Distance \\ from 0.5}} \\ \midrule
			\multicolumn{1}{c}{\multirow{2}{*}{\makecell[t]{Single-Pulse \\ + XOR}}} & \multicolumn{2}{c}{00-FF} & \multicolumn{2}{c}{FF-00} & \multicolumn{1}{c}{\multirow{2}{*}{0.5068}} & \multicolumn{1}{c}{\multirow{2}{*}{0.0068}} \\
			\multicolumn{1}{c}{} & \multicolumn{1}{c}{$\mu$ = 0.5987} & \multicolumn{1}{c}{$\sigma$=0.1405} & \multicolumn{1}{c}{$\mu$ = 0.5772} & \multicolumn{1}{c}{$\sigma$=0.1319} & \multicolumn{1}{c}{} & \multicolumn{1}{c}{} \\ 
			\multicolumn{1}{c}{\multirow{2}{*}{\makecell[t]{Dual-Pulse\\ + XOR}} } & \multicolumn{2}{c}{00-FF-00} & \multicolumn{2}{c}{FF-00-FF} & \multicolumn{1}{c}{\multirow{2}{*}{0.4990}} & \multicolumn{1}{c}{\multirow{2}{*}{0.0001}} \\ 
			\multicolumn{1}{c}{} & \multicolumn{1}{c}{$\mu$ = 0.5094} & \multicolumn{1}{c}{$\sigma$=0.0818} & \multicolumn{1}{c}{$\mu$ = 0.5040} & \multicolumn{1}{c}{$\sigma$=0.0819} & \multicolumn{1}{c}{} & \multicolumn{1}{c}{} \\ \bottomrule
		\end{tabular}%
	\end{table}

	\clearpage
	\section{Reconfigurability in 00-FF-00 polarized writing}
	
\begin{figure}[!th]
		\centering
		\includegraphics[width=0.5\linewidth]{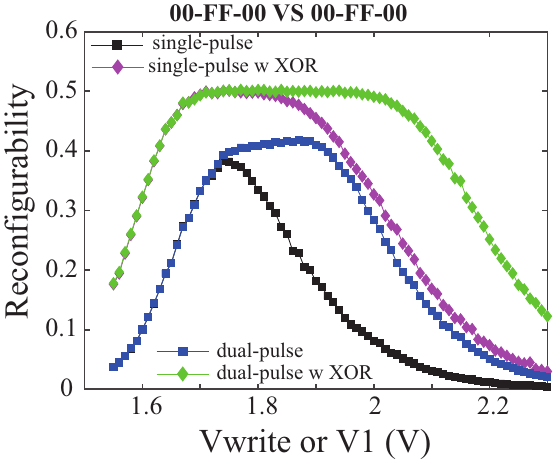}
		\caption{Reconfigurability (Inter-Reconfig. HD) as a function of write voltage amplitudes.}
		\label{figSI:Reconfig0}
	\end{figure}

	Reconfigurability is defined as Inter-Reconfig. HD. The 00-FF-00 polarized reconfiguration is shown in Fig. \ref{figSI:Reconfig0}, exhibiting a similar tendency to the FF-00-FF polarized reconfiguration. The combination of 00-FF-00 dual-pulse writing and XOR post-processing leads to unlimited reconfiguration counts and a large operation window.
	
\begin{figure}[!h]
		\centering
		\includegraphics[width=0.8\linewidth]{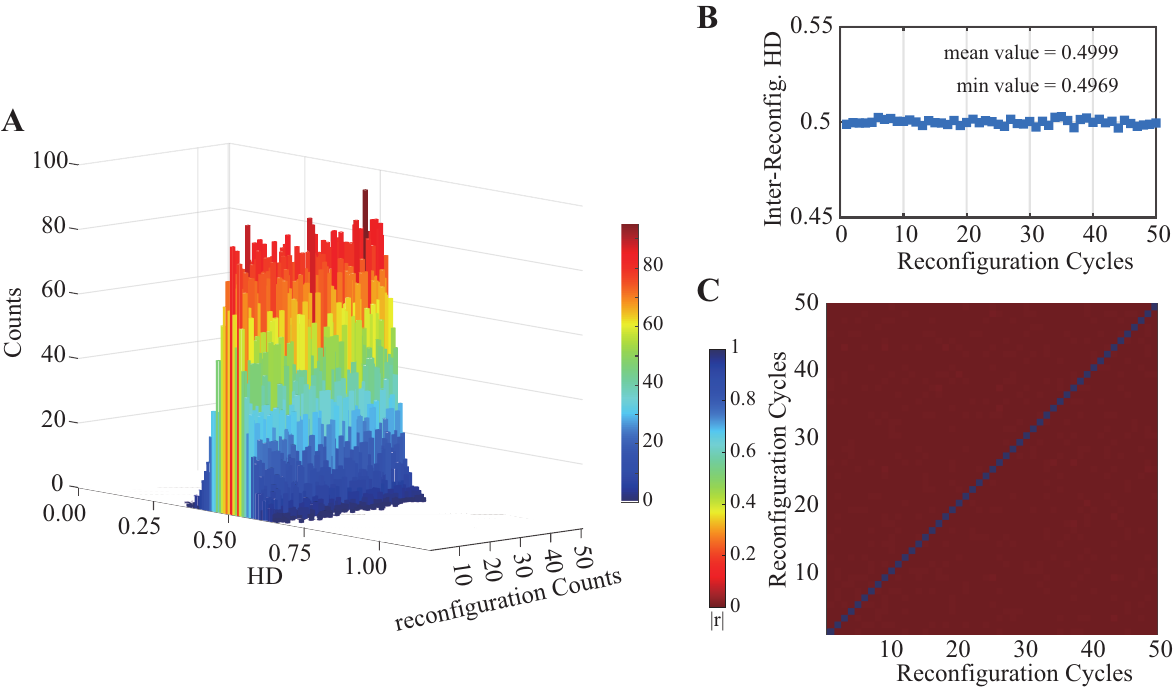}
		\caption{\textbf{Evaluation of reconfigurability considering 00-FF-00 polarized reconfigurations}. (\textbf{A}) Distribution of the normalized Inter-reconfig. HDs over 50 reconfigurations. (\textbf{B}) Statistics of the mean value of Inter-reconfig. HD. (\textbf{C}) Correlation matrix concerning the reconfiguration results.}
		\label{figSI:Reconfig002FF200}
	\end{figure}

	\clearpage
	Conducting over 50 counts of 00-FF-00 polarized reconfiguration, good reconfigurability is demonstrated in Fig. \ref{figSI:Reconfig002FF200}.
	
	\vspace{2cm}
	
	\clearpage
	\section{Temperature independence of SOT-track resistance}
	
	The material of the SOT track is mainly annealed $\beta$-phase tungsten ($\beta$-W). We tested the resistance of the SOT track in single SOT-MTJs, which are fabricated on the same platform but without CMOS. The resistance values of the SOT track at various temperatures are listed in Table \ref{TableSI: betaW}. The results indicate that the SOT track resistance of all samples exhibits negligible dependence on temperature. The minor fluctuations observed could be attributed to measurement error. Consequently, it is demonstrated that the resistivity of annealed $\beta$-W is essentially independent of temperature, consistent with the previous report\cite{hao2015beta}.
	
	\begin{table}[h]
		\caption{Resistance of SOT track measured under various temperatures}
		\centering
		\renewcommand{\arraystretch}{1.5} 
		\begin{tabular}{|p{0.2\textwidth}|c|c|c|c|c|}
			\hline
			\textbf{  } & \textbf{-40 ℃} & \textbf{   0 ℃} & \textbf{25 ℃} & \textbf{75 ℃} & \textbf{125 ℃} \\ \hline
			SOT-MTJ \#1 & 676 & 681 & 678 & 678 & 673 \\ \hline
			SOT-MTJ \#2 & 741 & 741 & 743 & 740 & 742 \\ \hline
			SOT-MTJ \#3 & 672 & 674 & 678 & 674 & 670 \\ \hline
			SOT-MTJ \#4 & 679 & 680 & 685 & 681 & 680 \\ \hline
			SOT-MTJ \#5 & 706 & 710 & 712 & 707 & 704 \\ \hline
			SOT-MTJ \#6 & 741 & 744 & 744 & 738 & 741 \\ \hline
			SOT-MTJ \#7 & 829 & 832 & 831 & 825 & 826 \\ \hline
			SOT-MTJ \#8 & 786 & 786 & 790 & 783 & 783 \\ \hline
			SOT-MTJ \#9 & 719 & 722 & 722 & 721 & 718 \\ \hline
			SOT-MTJ \#10 & 698 & 700 & 700 & 700 & 699 \\ \hline
			SOT-MTJ \#11 & 672 & 675 & 674 & 670 & 671 \\ \hline
			SOT-MTJ \#12 & 681 & 684 & 685 & 682 & 688 \\ \hline
			SOT-MTJ \#13 & 706 & 709 & 711 & 708 & 717 \\ \hline
			SOT-MTJ \#14 & 750 & 753 & 753 & 750 & 751 \\ \hline
		\end{tabular}
		\label{TableSI: betaW}
	\end{table}
	
	\clearpage
	\section{Reconfigurability under various working temperatures}
	
\begin{figure}[!h]
		\centering
		\includegraphics[width=0.9\linewidth]{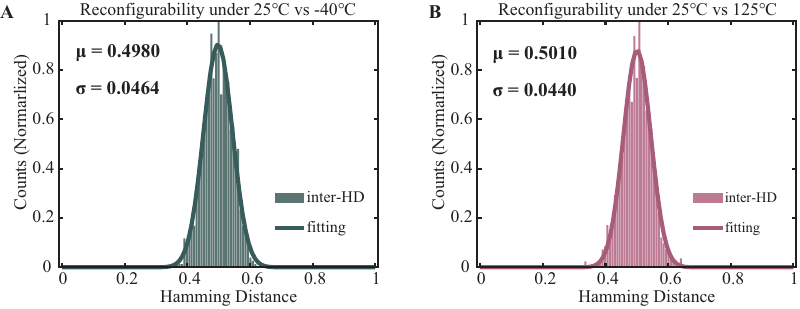}
		\caption{\textbf{Evaluation of Inter-Reconfig. HD}: (\textbf{A}) between the reconfiguration under 25 ℃ and that under -40 ℃, (\textbf{B}) between the reconfiguration under 25 ℃ and that under 125 ℃. }
		\label{figSI:Reconfig_Temp}
	\end{figure}

	Figure \ref{figSI:Reconfig_Temp}A shows the Inter-Reconfig. HD between the reconfiguration under 25 ℃ and that under 125 ℃. Figure \ref{figSI:Reconfig_Temp}B shows the Inter-Reconfig. HD between the reconfiguration under 25 ℃ and that under -40 ℃. Reconfigurability under various working temperatures proves the feasibility of unified reconfiguration settings. There is no need for sensor feedback or waiting for the environment to drop to room temperature. Therefore, the dual-pulse strategy increases the feasibility of operation and releases the need for a temperature feedback circuit module, showing excellent potential in hardware security.

	\clearpage
	\section{Demonstration of read reliability \label{SecReliability}}
	
	Bit Error Rate (BER) indicates the reliability of the data readout. The nominal conditions are defined as room temperature (25 ℃) and 1.8 V supply voltage, which refers to the supply voltage in the digital module that is mainly used for addressing, controlling the read transistor, etc. 
	
\begin{figure}[!h]
		\centering
		\includegraphics[width=0.5\linewidth]{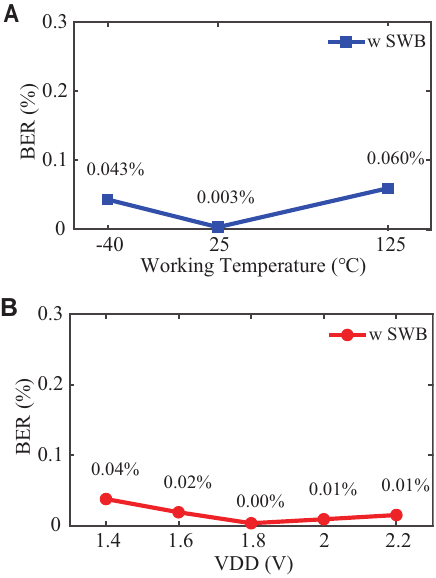}
		\caption{\textbf{Evaluation of reliability w SWB operation}. BER of the SOT-MRAM rPUF with (\textbf{A}) the temperature ranging from -40 ℃ to 125 ℃, and (\textbf{B}) the supply VDD voltage ranging from 1.4 V to 2.2 V.}
		\label{figSI-reliability-SWB}
	\end{figure}

\begin{figure}[!h]
		\centering
		\includegraphics[width=0.5\linewidth]{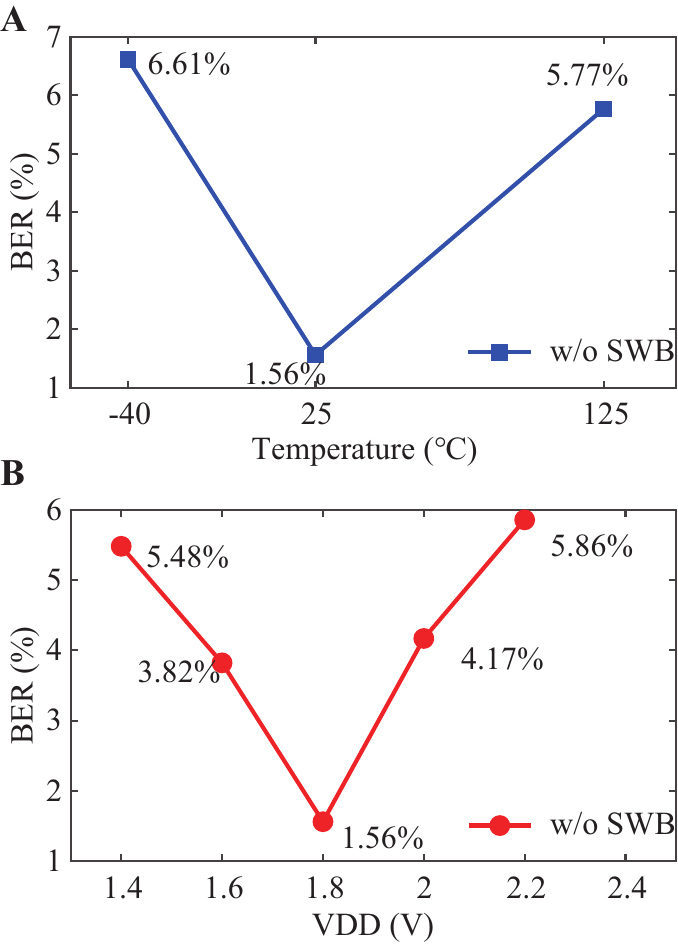}
		\caption{\textbf{Evaluation of reliability w/o SWB operation}. BER of the SOT-MRAM rPUF with (\textbf{A}) the temperature ranging from -40 ℃ to 125 ℃, and (\textbf{B}) the supply VDD voltage ranging from 1.4 V to 2.2 V.}
		\label{figSI-reliability}
	\end{figure}

	Figure \ref{figSI-reliability-SWB} and \ref{figSI-reliability} present the BER result with and without SWB, respectively, and all data are based on the average of 15 times of readout. With the help of SWB, the BER under the nominal condition is estimated to be $3.29 \times 10^{-5}$, $\sim$474 $\times$ improved reliability compared to the results w/o SWB. In addition, PUF with SWB also shows good read reliability in the cases of varying temperature and supply voltage.
	
	\clearpage
	\section{Issues reflected by slope $k$\label{SecParameters}}
	
	The slope $k$ reflects the discrete degree of device performance; meanwhile, the slope $k$ is the essential parameter to solve $\beta$.
	
	We perform numerical modeling, with preliminary assumptions of introducing the process variation of the SOT track width. The numerical modeling follows the steps below:
	
	(i) Setting baseline case and calculating the initial critical current $J_{c0}$.
	
	(ii) Introducing bottom electrode width variation. Generate the SOT track width for each SOT-MTJ through a random distribution by introducing the coefficient of variation (CV), which is expressed as $\sigma/\mu$. 
	
	(iii) Device-level switching probability curves. For each SOT-MTJ, the switching probability curve as a function of current Psw(V) differs in various SOT track widths. 
	
	(iv) Signal-driven switching probability and data generation. Based on the input write signals (e.g., specific current/voltage), query the corresponding switching probability, and generate the storage bits by combining with the uniform distribution. 
	
	(v) Hamming weight Statistics. Iterate through all SOT-MTJs and count the percentage of successful switching. 
	
	(vi) Multi-signal scanning and curve generation. Vary the write signal parameters (e.g., voltage step size) and repeat steps iv-v to generate the curve of Hamming Weight with the signal. 
	
	(vii) Slope k calculation. At Hamming weight = 50\%, a linear fit to the curve was performed and the slope k was extracted.
	
	\begin{figure}[!th]
		\centering
		\includegraphics[width=0.8\linewidth]{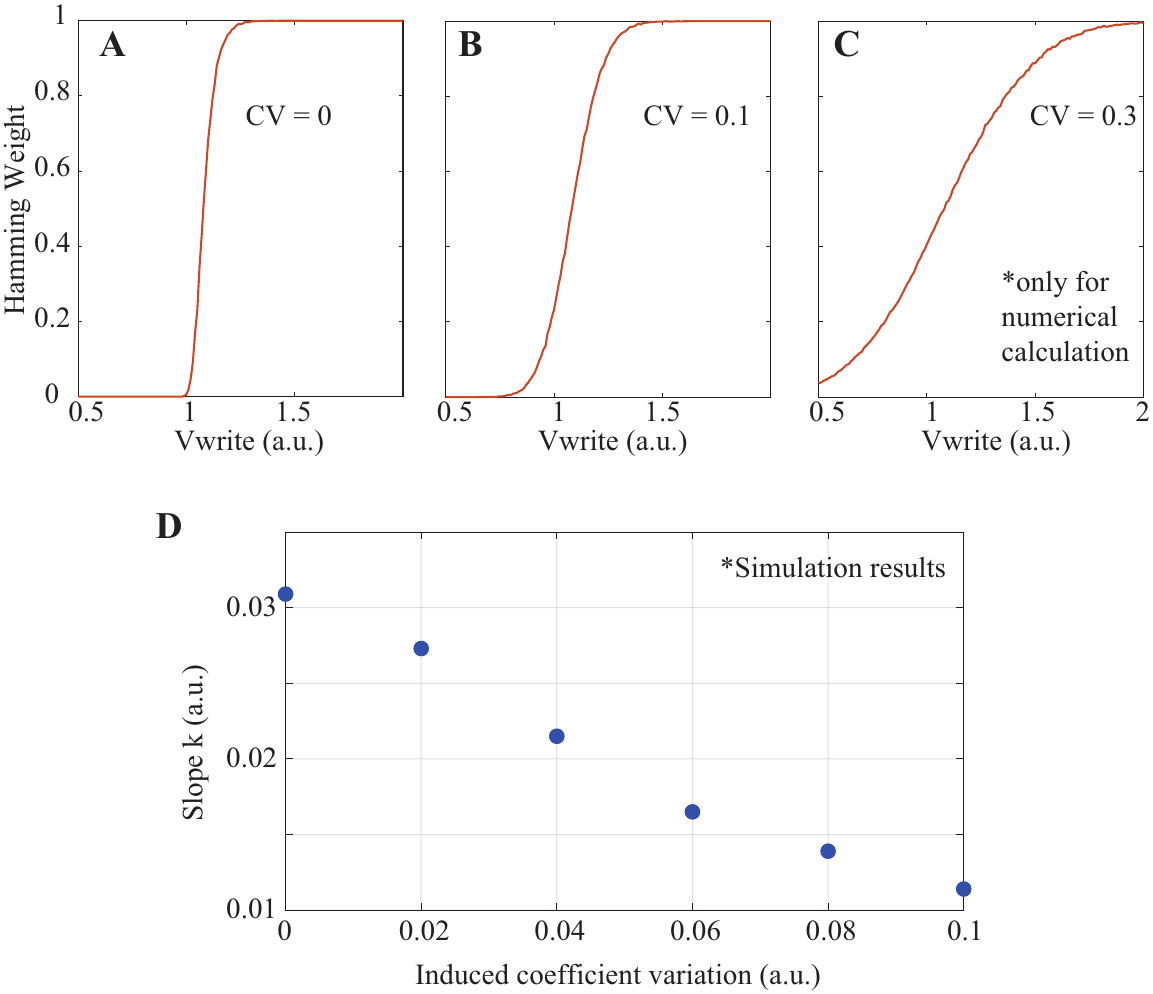}
		\caption{\textbf{Numerical prediction of slope k}. Hamming weight as a function of applied voltage in the case of (\textbf{A}) no process variation, (\textbf{B}) with coefficient variation of 0.1, (\textbf{C}) coefficient variation of 0.3. The case of (A) equals to the switching probability of the single SOT-MTJ. (\textbf{D}) The evaluated k as a function of process variation. The induced coefficient variations are set only for numerical calculation.}
		\label{figSI-slopek}
	\end{figure}
	
	The parameter k is a function of variation and the MTJ magnetic parameters. When the preparation is totally ideal and there are no process deviations, all SOT-MTJs exhibit consistent properties. At this point, the HW curve should coincide with the Psw curve of a single SOT-MTJ, as shown in Fig. \ref{figSI-slopek}A. This situation is not possible in practice. When process bias exists, the HW curve aggregates the situation of all SOT-MTJs in the array. When the process deviation is larger, the difference between SOT-MTJs becomes larger and the HW curve rises more slowly, as shown in Fig. \ref{figSI-slopek}B-C. For the slope at 50\% is extracted and fitted, as shown in Fig. \ref{figSI-slopek}D. 
	
	
	The model indicates that $k$ reflects the process variation. It should be noted that Fig. \ref{figSI-slopek} only explains the influence of process variation, where the extracted values are not the experimental results.
	
	\clearpage
	\section{Explanation of the dual-Pulse modeling\label{SecDualPulse}}
	
	\subsection{Independent events in the case of the 00-FF and FF-00 pulses\label{SecIndep}}
	
	\begin{figure}[!h]
		\centering
		\includegraphics[width=0.4\linewidth]{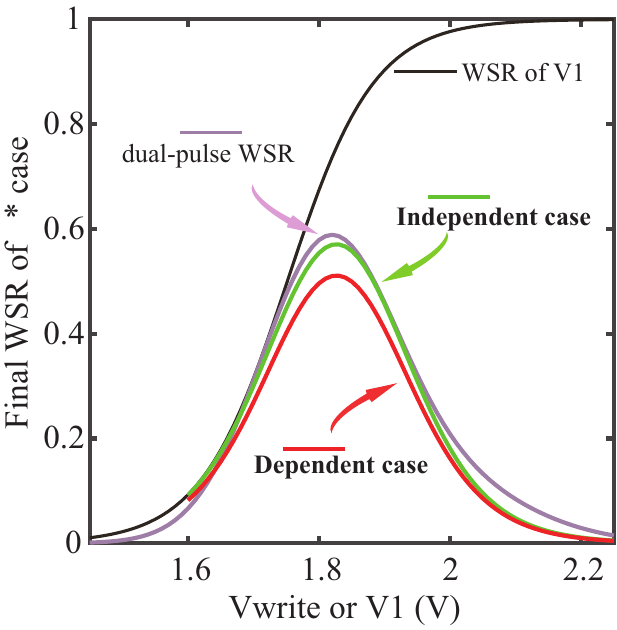}
		\caption{\textbf{Verification of dual-pulse modeling.} The experimental results align well with the simulated results in dual-pulse reconfiguration. Independent case means Eq. \ref{eqSI:independent}. Dependence case refers to Eq. \ref{eqSI:dependent}.}
		\label{figSI-independent}
	\end{figure}
	
	Figure \ref{figSI-independent} confirms that the bit switching behaviors in the first and second pulses are independent events. From the unit perspective: assuming that the MTJ holds the switching probability $P_1$ ($P_2$) of 00-FF (FF-00) polarized writing when applying only one pulse, the final time-domain distribution probability is expressed as $P_1 - P_1 \times P_2$ when applying 00-FF-00 polarized dual-pulse writing. 
	
	From the array perspective, the WSR shows the array-level statistic results in the spatial-domain distribution, reflecting the whole behavior of all single units. Assuming that the array holds the $WSR1$ ($WSR2$) of 00-FF (FF-00) polarized single-pulse writing, the final WSR (noted as $F$) after applying dual pulses is given by:
	
	\begin{equation}
	F = WSR1 - WSR1 \times WSR2
	\label{eqSI:independent}
	\end{equation}
	
	The modeling in Eq. \ref{eqSI:independent} requires two premises: 1. unit switching processes are independent events; 2. The distribution of switching probability holds randomness. 
	
	The following is the case of the dependent event. If $WSR1$ and $WSR2$ hold a dependent relationship in domain distribution, the final WSR ($F$) would be far smaller than the predicted value. For example, if the units switched by the second pulse are a subset of those affected by the first pulse, the relationship can be expressed as follows:
	\begin{equation}
	F = WSR1 - WSR2
	\label{eqSI:dependent}
	\end{equation}
	
	As shown in Fig. \ref{figSI-independent}, results of Eq. \ref{eqSI:independent} show a better agreement compared with that of Eq. \ref{eqSI:dependent}. And the mean absolute error for the model of Eq. \ref{eqSI:independent} and experimental is about 0.02, which confirms again the strong independent event according to the first and the second pulse writing results.
	
	\subsection{Calculation of $\beta$ in simplified model}
	
	As the verification of the independent case in Sec. \ref{SecIndep}, the final WSR \( F \) after two write operations has an algebraic relationship with \( WSR1 \) and \( WSR2 \), given by 
	\begin{equation}
	F(V1,V2) = WSR1(V1) - WSR1(V1) \cdot WSR2(V2)
	\end{equation}
	
	In this subsection, we first makes a simplification of the asymmetry of the two pulses. \uline{\textbf{Assumption (a)}}: Pulses of different polarities have approximate WSRs.
	\begin{equation}
	WSR1 (V) = WSR2 (V)
	\end{equation}
	
	In this case,
	\begin{equation}
	F(V1,V2) = WSR1(V1) - WSR1(V1) \cdot WSR1(V2)
	\end{equation}
	
	Since the sigmoid-like Psw function grows sharply around 50\%, a local fit can be done with a tangent line to form a unitary function. \uline{\textbf{Assumption (b)}}: WSR1 could be simplified to a tangent line, which is only valid around the critical VSOT.
	\begin{equation}
	WSR1 (V) = k\times V +b
	\end{equation}
	
	Because of \(|V_2| = |V_1| - \beta \), \( F \) can be expressed as
	\begin{align}
	F(V1,V2) &= WSR1(V1) - WSR1(V1) \cdot WSR1(V1 - \beta) \\
	& = WSR1(V1) - WSR1(V1) \cdot (WSR1(V1) - k\beta)
	\end{align}

	then, 
	\begin{align}
	F(V1,V2) & = (1 + k\beta)\cdot WSR1(V1) - WSR1(V1)^2 \\
	& = -k^2 V1^2 + (k+k^2\beta-2kb)\cdot V1 + (b+k\beta b-b^2)
	\end{align}
	
	To solve for the value of $\beta$, the extreme value could be resolved as:
	
	\begin{equation}
	F_{extreme} = \frac{1}{4}(1 + k \beta)^2 
	\end{equation}
	
	Here we define the available value of $\beta$ are those values making $F_{extreme}$ within the target window, for a continuous and large operating window of available voltage, which also corresponds to the region in trend (ii) in the main manuscript. A target window of (0.4, 0.6] is used, which is the result without the XOR operation, and the post-processing would adjust the results within this window to a more ideal range. By solving \(0.4< F_{extreme} \leq 0.6\), 
	
	\begin{equation}
	\beta \in [-0.6829, -0.6068) \cup (0.0710, 0.1471]
	\end{equation}
	
	Since Assumption (b) is only valid around the critical VSOT, the solution of [-0.6829, -0.6068) is discarded, and (0.0710, 0.1471] is valid.
	
	Thus, the available $\beta$ is  (0.0710, 0.1471]. 
	
	\begin{equation}
	\beta \in (0.0710, 0.1471]
	\end{equation}
	
	The optimized case means that the extreme value is equal to the boundary of the target window.
	\begin{equation}
	F_{extreme} = Upperbound
	\end{equation}
	
	Therefore, 
	\begin{equation}
	\boxed{Optimal~\beta = 0.147~V}
	\end{equation}

	\subsection{Extended modeling}
	If we also consider asymmetries, offsets, etc., in our chip, we can modify the simplified model of the previous subsection, with different slopes and intercepts,
	\begin{equation}
	WSR1 (V) =k_1\times V + b_1
	\end{equation}
	
	\begin{equation}
	WSR2 (V)=k_2\times V + b_2
	\end{equation}
	
	\( F \) is expressed as \( WSR1(|V_1|) - WSR1(|V_1|) \cdot WSR2(|V_1|-\beta) \). The rest of the computational procedure is similar to the previous subsection.
	

	
	\clearpage
	\section{Batch demonstration of the selected $\beta$\label{SecBatch}}
	
	$\beta$ is demonstrated to be a stable parameter for SOT-MRAM chips within the same batch. Table \ref{TableSI:10chips} lists wide-temperature-range measured results of 10 chips, all of which show the temperature-resilient reconfiguration. This finding facilitates the chip design and reconfiguration operation. 
	
	\begin{table}[h]
		\caption{Verification of reconfigurability by applying unified operations in the same batch of SOT-MRAM chips} 
		\centering
		\renewcommand{\arraystretch}{1.5}  
		\begin{tabular}{|p{0.13\textwidth}|p{0.2\textwidth}|p{0.2\textwidth}|p{0.3\textwidth}|}
			\hline
			\makecell{No. of \\ samples} & \makecell{ HW \\ @ -40 ℃} & \makecell{ HW \\ @ 125 ℃}  & \makecell{Reconfigurability}\\ \hline
			\#1 &0.50014 	&0.50041 	&0.49607  \\ \hline
			\#2 &0.49709 	&0.50224 	&0.50137  \\ \hline
			\#3 &0.49647	&0.50275	&0.49995  \\ \hline
			\#4 &0.49935 	&0.50413 	&0.49832  \\ \hline
			\#5 &0.49803 	&0.49905 	&0.50198  \\ \hline
			\#6 &0.49876 	&0.50076 	&0.49973  \\ \hline
			\#7 &0.49874 	&0.49651 	&0.49992  \\ \hline
			\#8 &0.49934 	&0.49882 	&0.49988  \\ \hline
			\#9 &0.49886 	&0.49990 	&0.49933  \\ \hline
			\#10 & 0.50423 	&0.49822 	&0.49809  \\ \hline
		\end{tabular}
		\label{TableSI:10chips}
	\end{table}

	\clearpage
	\section{Chip architecture optimization perspective}
	
	\begin{figure}[!h]
		\centering
		\includegraphics[width=0.5\linewidth]{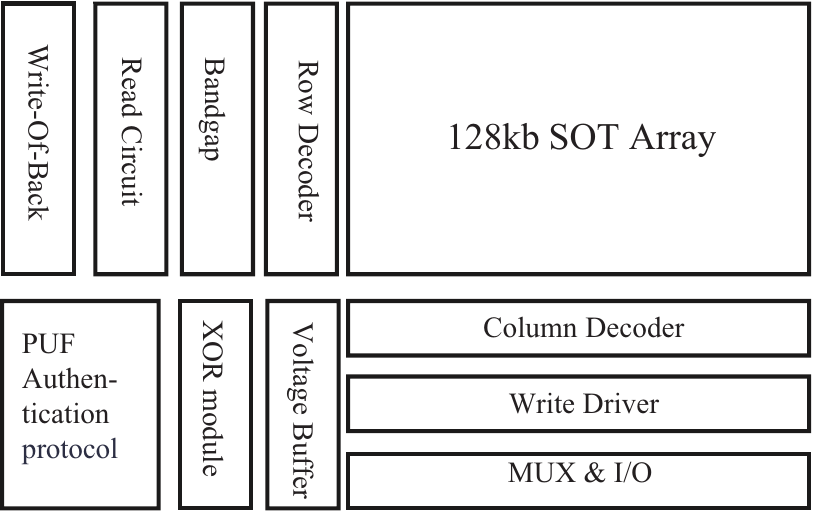}
		\caption{Chip structure as a perspective.}
		\label{figSI:structure}
	\end{figure}
	
	As a prospect, our proposal is feasible and holds potential to design a multi-mode chip, with a structure shown in Fig. \ref{figSI:structure}. Overall, the integrated chip contains the PUF authentication protocol, a post-processing module, the bandgap circuit, the reference buffer circuit, the read/write circuit, the row/column decoder, I/O, etc. The bandgap circuit can produce a stable reference voltage that does not vary with temperature. The reference buffer circuit outputs a stabilized reference voltage to the write circuit.
	
	Specifically, the amplitude of $V1$ and $\beta$ can be obtained from the built-in memory of the PUF or deduced from empirical parameters. After that, the reference voltage buffer circuit processes the amplitude of the first pulse and the amplitude of the second pulse based on the reference voltage and outputs two write voltages. Finally, reconfiguration of the PUF is realized after inputting two write voltages to the write driver.
	
	\bibliography{reference} 
	\bibliographystyle{sciencemag}